\documentclass[10pt,twocolumn]{article}
\usepackage[utf8]{inputenc}
\usepackage[T1]{fontenc}
\usepackage{lmodern}
\usepackage[a4paper,left=1.5cm,right=1.5cm,top=1.9cm,bottom=2.0cm,columnsep=0.6cm]{geometry}
\usepackage{amsmath,amssymb}
\usepackage{graphicx}
\usepackage{bm}
\usepackage{booktabs}
\usepackage[numbers,sort&compress]{natbib}
\usepackage{titlesec}
\usepackage{flafter}   
\usepackage[font=small,labelfont=bf,labelsep=period,justification=justified,skip=4pt]{caption}
\usepackage[colorlinks=true,allcolors=blue,bookmarks=false]{hyperref}

\titleformat{\section}{\normalfont\normalsize\bfseries}{\thesection.}{0.5em}{}
\titlespacing*{\section}{0pt}{1.4ex plus 1.2ex minus 0.2ex}{0.5ex plus 0.3ex}
\AtBeginDocument{%
  \setlength{\abovedisplayskip}{5pt plus 4pt minus 2pt}\setlength{\belowdisplayskip}{5pt plus 4pt minus 2pt}%
  \setlength{\abovedisplayshortskip}{3pt plus 2pt}\setlength{\belowdisplayshortskip}{4pt plus 2pt minus 1pt}}
\begin{document}

\twocolumn[
\begin{center}
{\Large\bfseries Tunable Microresonator Arrays Induced by Geometric Phase\par}
\vspace{1.6ex}
{\large M. Sumetsky\par}
\vspace{0.6ex}
{\small\textit{Aston Institute of Photonic Technologies, Aston University, Birmingham B4 7ET, UK}\\ m.sumetsky@aston.ac.uk\par}
\vspace{1.6ex}
\begin{minipage}{0.86\textwidth}\small
Unlike periodic field-intensity patterns, which can be generated by interference of coherent beams of uniform intensity, periodic structures in a material are usually created by periodic modification of its properties or geometry. In contrast, we show here that a millimeter-long bundle of four conventional side-coupled optical fibers with a slowly varying transverse arrangement can support a nearly periodic array of whispering-gallery-mode microresonators, whose period can be tuned mechanically down to about 100~$\mu$m.
\end{minipage}
\end{center}
\vspace{24pt}  
]

\section*{Introduction}

Periodic optical structures are used as wavelength-selective reflectors, filters and sensors~\cite{r1}, as photonic band-gap waveguides and cavities~\cite{r2}, and as slow-light delay lines~\cite{r3,r4}. Their period is written into a material, as in fiber Bragg gratings~\cite{r1} and chiral fiber gratings~\cite{r5,r6}, fabricated as a periodic dielectric structure~\cite{r2}, built as an array of identical waveguides or of coupled microresonators~\cite{r3,r7,r8}, including chains of microresonators fabricated along an optical fiber~\cite{r9}, or induced by a periodic field, such as an acoustic wave~\cite{r10,r11} or the intensity pattern of interfering beams in a nonlinear medium~\cite{r12}. A linear refractive-index gradient across a waveguide array makes a light beam oscillate periodically along the array (Bloch oscillations), but the array itself remains uniform along the propagation direction~\cite{r13}.

\begin{figure}[t]
\centering\includegraphics[width=\columnwidth]{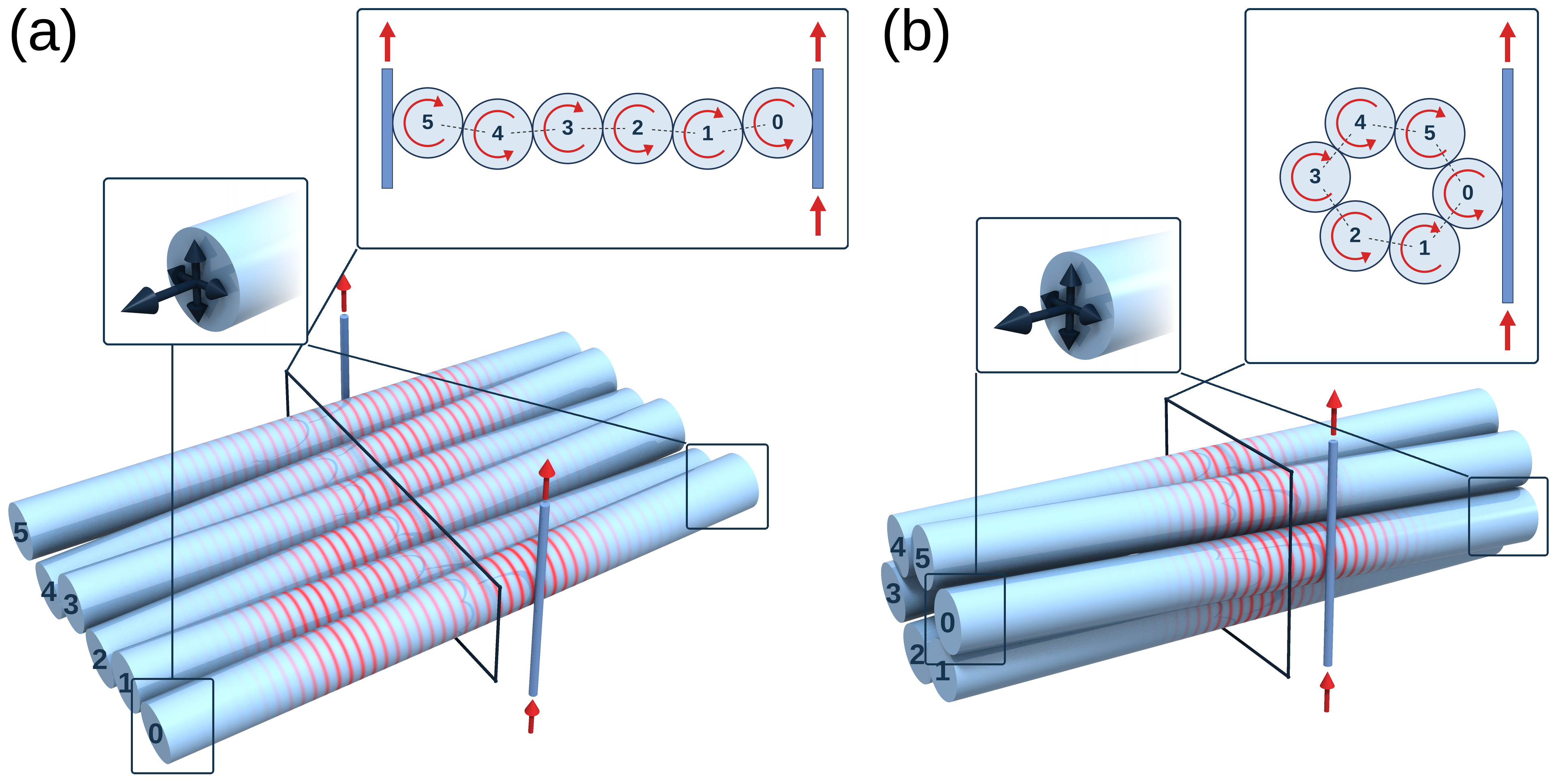}
\caption{\label{fig1}Arrangements of side-coupled optical fibers. (a) Quasi-planar stack of slightly tilted fibers, which form an open chain. (b) Three-dimensional bundle, in which the fibers close a coupling loop. Red lines on the fibers show WGMs excited through a transverse microfiber, and red arrows show the input and output light. The rectangle marks the cross-section shown at the upper right, with the fibers numbered and the directions of the WGMs, which are opposite in neighboring fibers. In the cross-section insets, possible interfiber gaps are not shown. Upper left insets: each fiber end can be translated transversely and stretched.}
\end{figure}

However, in every case considered, the period of the structure equals the period of the fabrication pattern or of the intensity of the applied field. To our knowledge, a way to create a \emph{periodic potential that localizes light within periods much shorter than the length over which the material properties change} has not been envisioned to date. Here we show that such a potential can be created by a smooth change of the transverse geometry of an optical fiber bundle. The period is set by the \emph{geometric phase} of the bundle which does not repeat along the fiber length.

The geometric phase appears when light can go from one waveguide or resonator to another along two or more paths forming a closed loop. This loop phase, known as the Aharonov--Bohm phase~\cite{r14} and, more generally, as a geometric phase~\cite{r15,r16,r17,r18}, shifts the eigenfrequencies of the coupled system. It provides effective magnetic fields for photons~\cite{r19,r20,r21,r22,r23} and localizes light in Aharonov--Bohm cages~\cite{r24,r25,r26}. When generated by twisting, it inhibits tunneling between the cores of multicore fibers~\cite{r27,r28} and controls light in rotated waveguides~\cite{r29} and microrings~\cite{r30}. In all these systems, the optical field varies along the structure on the same length scale as the structure itself.

For the potential to have a period much shorter than the length over which the material geometry changes, the loop phase $\Phi$ must change by $2\pi$ when the material geometry changes by only a small fraction of its own size. Remarkably, \emph{whispering-gallery modes (WGMs) slowly propagating along optical fibers} provide such a phase. At wavelength $\lambda\sim 1.5$~$\mu$m, a WGM circulating around a fiber or a resonator of radius $r_{0} \sim 20$--$100$~$\mu$m and refractive index $n_{0}\sim 1.5$~\cite{r31,r32,r33} has the azimuthal number
\begin{equation}\label{eq1}
m\approx\frac{2\pi n_{0}r_{0}}{\lambda}\approx 10^{2}-10^{3}.
\end{equation}

When two touching fibers are tilted by a small angle $\alpha$, the phase of their coupling changes along the fibers at the rate $\beta_{0} = m\alpha/r_{0}$~\cite{r34}, so a rotation of the contact points by an angle of the order of $1/m$, a fraction of a degree, changes the loop phase by $2\pi$. Near its cutoff, a WGM obeys a Schrödinger-type equation for the axial envelope, in which the cutoff frequency is the potential~\cite{r35}. This is the basis of surface nanoscale axial photonics (SNAP), where nanoscale radius variations form microresonators~\cite{r9,r35,r36}.

Potential wells that localize WGMs and are tuned mechanically have been demonstrated with a single and two touching fibers, by bending~\cite{r37}, by tilt and translation~\cite{r34,r38,r39,r40} and by stretching~\cite{r41}. Generalization of these configurations to multiple coupled fiber systems illustrated in Fig.~\ref{fig1}(a) is a promising way of creating miniature photonics circuits with small loss (characteristic of silica fibers), reconfigurability (enabled by translation and stretching of the fiber ends, illustrated by the insets) and precision (enabled by the exceptional uniformity of optical fibers and accuracy of nanopositioners). However, the quasi-planar arrangement of fibers shown in Fig.~\ref{fig1}(a) forms an open chain, which has no loop phase. The potential supporting microresonators is then formed by the variation of the coupling magnitude along the fiber axis, while the phase of the coupling, which grows linearly along the fibers~\cite{r34}, can modify the eigenmodes of this microresonator and does not make the structure periodic (Supplemental Material (SM), S1).

Alternatively, a bundle of more than two fibers can have a closed coupling loop as illustrated in Fig.~\ref{fig1}(b). We show below that in the bundle of Fig.~\ref{fig1}(b) the geometric phase of WGMs creates a lattice of axial potential wells. As shown below, for $N$ side-coupled fibers with arbitrary cutoff frequencies and equal coupling magnitudes, the local cutoff frequencies depend on the coupling phases only through their sum $\Phi$ around the loop. This sum is determined by the cross-sectional configuration of the bundle and does not change under a rigid twist, in contrast with twisted multicore fibers~\cite{r27}. A shape-dependent $\Phi$ exists only for an \emph{even number of fibers}, so four fibers are the smallest such bundle. For the four-fiber example considered below, whose cross-section changes linearly from a rhombus through a square to the rotated rhombus (Fig.~\ref{fig2}), the local WGM cutoff frequencies are periodic functions of the axial coordinate, and each change of the rhombus angle by about $1^{\circ}$ gives one period of this axial potential. The bundle then carries four interleaved lattices of microresonators with a period of the order of 100~$\mu$m and a depth of about 80 GHz, which is equivalent to an effective fiber radius variation of about 26 nm.

\section*{Model}

\begin{figure}[t]
\centering\includegraphics[width=\columnwidth]{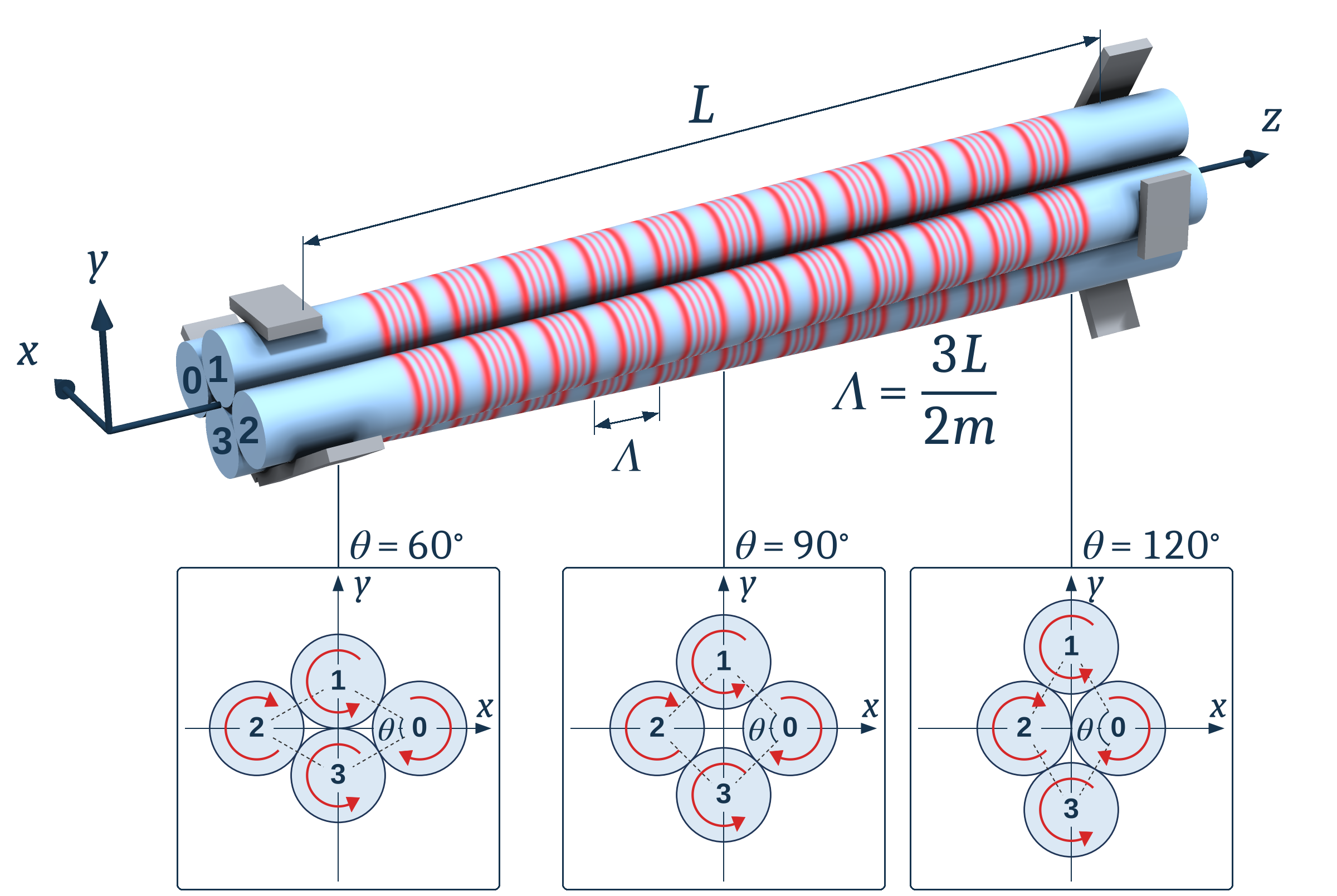}
\caption{\label{fig2}Four side-coupled fibers sheared from a rhombus (left), through a square (center), to the same rhombus rotated by $90^{\circ}$ (right). Top: schematic bundle; the red lines show WGMs localized in a periodic array of microresonators. $L$ is the length of the bundle between the holders. $\Lambda$ is the distance between neighboring microresonators. For a bundle sheared over the whole window from $60^{\circ}$ to $120^{\circ}$, as shown, $\Lambda=3L/(2m)$, where $m$ is the azimuthal number of the WGM. The $z$ axis is directed along the bundle center, and the $x$ and $y$ axes along the diagonals of the rhombus formed by the fiber centers. Bottom: cross-sections at the left end, the center and the right end, viewed from the right end, against the $z$ axis. Fibers 0 and 2 lie on the $x$ axis, and fibers 1 and 3 on the $y$ axis. Red arrows show the WGMs, which circulate in opposite directions in neighboring fibers.}
\end{figure}

We start with consideration of a \emph{four-fiber model}. First, we assume that fibers are identical, have radius $r_{0}$ and are in direct side contact as illustrated in Fig.~\ref{fig2}. We direct the $z$ axis along the bundle center and the $x$ and $y$ axes along the diagonals of the rhombus formed by the fiber centers. In each transverse plane the centers of fibers $j = 0,1,2,3$ lie at $(x,y) = (a,0)$, $(0,b)$, $( - a,0)$, $(0, - b)$, where $a = 2r_{0}\cos(\theta/2)$ and $b = 2r_{0}\sin(\theta/2)$ are the half-diagonals, so every side of the rhombus has the length $2r_{0}$ and $\theta$ is its interior angle at fibers 0 and 2. For simplicity, we also assume that the angle varies linearly with $z$, $\theta(z) = \pi/2 + \sigma z$, where $z = 0$ at the central square cross-section and $\sigma$ is the gradient of the rhombus angle. This is the profile that equally tensioned fibers adopt between two holders (see \textit{Possible realizations} below and SM, S6). The side contacts persist for $60^{\circ} \leq \theta \leq 120^{\circ}$, as at these limits the two fibers on the shorter diagonal of the rhombus come into contact.

Near cutoff, the axial propagation constant of a WGM with azimuthal number $m$ is given by $\beta^{2} = \left( \omega - \omega_{c} \right)/D$~\cite{r34,r35}, where $\omega$ is the angular frequency of light, $\omega_{c}$ is the cutoff frequency of the WGM, and the constant $D$ is given in Eq.~(\ref{eq2}) below. The coupling coefficient of two touching fibers tilted by $\alpha$ has the phase $\beta_{0}z \approx m\alpha z/r_{0}$~\cite{r34}, which changes along the fiber axis $m$ times faster than angle $\theta$. The WGMs of touching fibers $j$ and $j + 1$ circulate in opposite directions, so their azimuthal numbers have opposite signs and alternate around the loop, $m_{j} = - m_{j + 1} = {( - 1)}^{j}m$. Each side of the rhombus then acquires the same phase $\phi(z) = m\left[ \theta(z) - \pi/2 \right]$ (SM, S3), and the envelopes $\psi_{j}(z)$ of the WGMs in the fibers obey
\begin{equation}\label{eq2}
\begin{aligned}
D\frac{d^{2}\psi_{j}}{dz^{2}}&+\left(\omega-\omega_{c}\right)\psi_{j}-Je^{i\phi(z)}\psi_{j+1}-Je^{-i\phi(z)}\psi_{j-1}=0,\\
D&=\frac{c^{2}}{2n_{0}^{2}\omega_{c}},\qquad (j\ \mathrm{mod}\ 4),
\end{aligned}
\end{equation}
where $c$ is the speed of light in vacuum, $n_{0}$ is the refractive index of the fiber, and $J$ is the coupling magnitude, which is constant because adjacent fibers remain in direct contact along the whole bundle (see \textit{Possible realizations} below). The phase of each envelope $\psi_{j}(z)$ is defined up to a constant, which is set by the choice of the origin of the azimuthal angle in fiber $j$. The substitution $\psi_{j}(z) \rightarrow e^{i\eta_{j}}\psi_{j}(z)$ with constants $\eta_{j}$ changes the phase of the coupling between fibers $j$ and $j + 1$ by $\eta_{j + 1} - \eta_{j}$ and does not change the sum of the coupling phases around the loop, which is the loop phase
\begin{equation}\label{eq3}
\Phi(z)=4\phi(z)=4m\left[\theta(z)-\pi/2\right]\quad(\mathrm{mod}\ 2\pi).
\end{equation}

The loop phase $\Phi(z)$ changes by $2\pi$, and the phase $\phi(z)$ of each coupling by $\pi/2$, when the rhombus angle $\theta(z)$ changes by $\pi/(2m)$, which is about $0.25^{\circ}$ for $m \approx 350$ of a 62.5~$\mu$m radius silica fiber at the wavelength $\lambda\approx 1.55$~$\mu$m. The loop phase of a bundle of $N$ fibers is considered below.

\section*{Microresonator lattice}

Eq.~(\ref{eq2}) can be separated into four independent equations for the combinations $\chi_{{\ell}}(z) = \frac{1}{2}\sum_{j = 0}^{3}e^{- i\pi{\ell}j/2}\psi_{j}(z)$, $\ell=0,1,2,3$, whose coefficients do not depend on $z$. The combination $\chi_{{\ell}}(z)$ is the amplitude of the \emph{supermode ${\ell}$} of the four fibers, in which all fibers carry the same intensity and the phase changes by $\pi\ell/2$ from one fiber to the next. Each supermode obeys its own Schrödinger-type equation,
\begin{equation}\label{eq4}
D\frac{d^{2}\chi_{\ell}}{dz^{2}}+\left[\omega-\omega_{c}-2J\cos\left(\frac{\pi\ell}{2}+\phi(z)\right)\right]\chi_{\ell}=0,
\end{equation}
in which $\omega_{c} + 2J\cos\left( \frac{\pi{\ell}}{2} + \phi(z) \right)$ is the local cutoff frequency of the supermode and plays the role of the potential. For the linear shear, $\phi = Kz$ with $K = m\sigma$, Eq.~(\ref{eq4}) is the Mathieu equation~\cite{r42} and the potential has the period $4\Lambda$, where
\begin{equation}\label{eq5}
\Lambda=\frac{\pi}{2m|\sigma|},
\end{equation}
is the distance, independent of $J$, over which the loop phase $\Phi$ changes by $2\pi$ and the phase $\phi$ of each coupling by $\pi/2$. For a conventional optical fiber with radius $r_{0} = 62.5$~$\mu$m, the distance $\Lambda$ corresponds to a change of the rhombus angle by $\pi/(2m) \approx 0.0045$ rad.

\begin{figure}[t]
\centering\includegraphics[width=0.9\columnwidth]{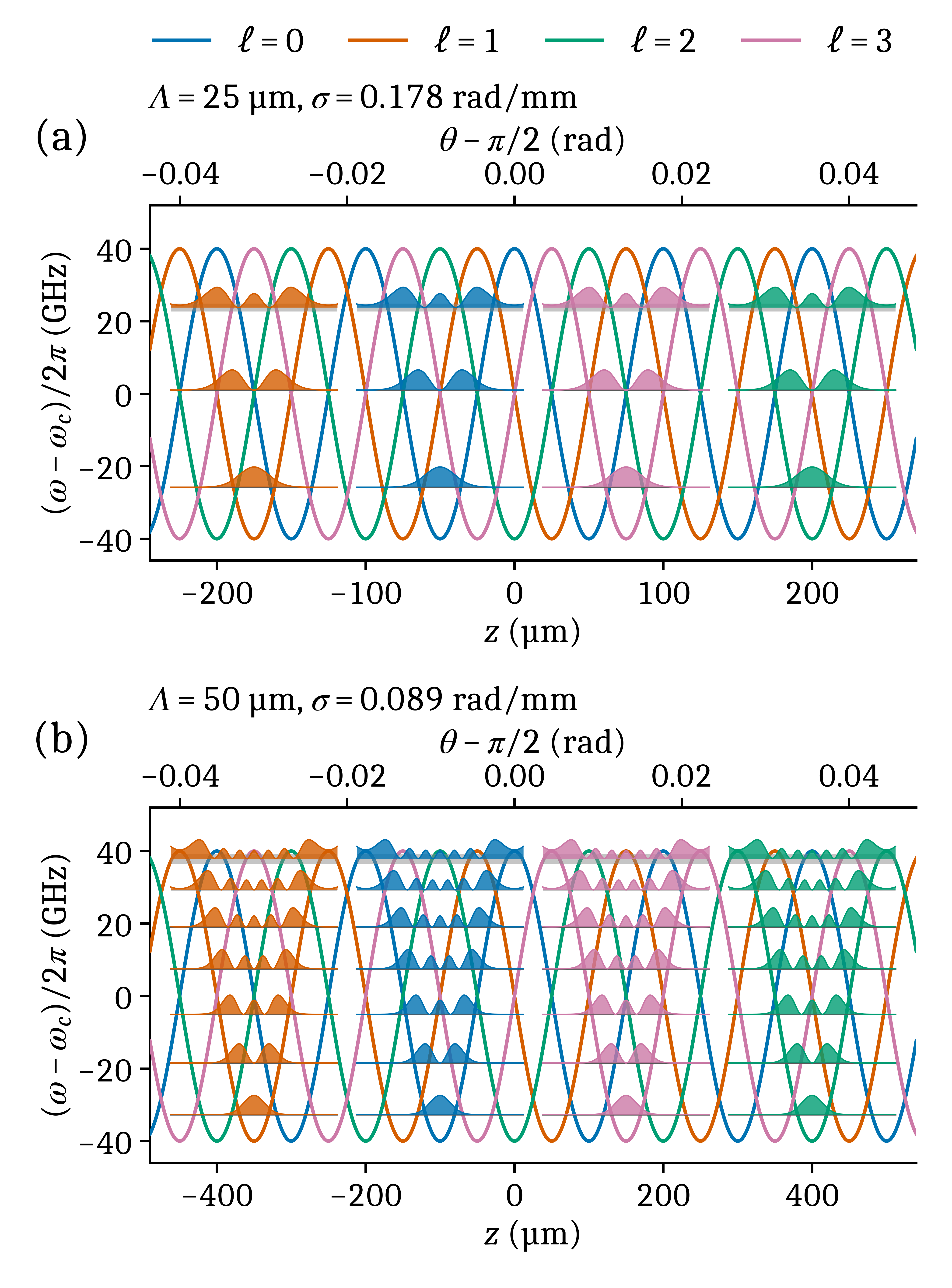}
\caption{\label{fig3}Lattices of microresonators in a bundle of four identical fibers for (a) $\Lambda = 25$~$\mu$m and (b) $\Lambda = 50$~$\mu$m; $J/2\pi = 20$ GHz, $m = 353$. Curves: local cutoff frequencies of the four supermodes, Eq.~(\ref{eq4}), which have the period $4\Lambda$. Horizontal lines: eigenfrequencies (gray strips where the band is wider than 0.3 GHz). Shaded: axial intensity of the modes at each eigenfrequency. The four supermodes have the same eigenfrequencies and modes in every potential well; for clarity the modes of each supermode are shown in one well only. Upper axes: rhombus angle relative to the square.}
\end{figure}

The four supermodes are supported by identical lattices of period $4\Lambda$ and depth $4J$, displaced by $\Lambda$ from one another (Fig.~\ref{fig3}). Their cutoff frequencies cross at $\omega_{c}$ and at $\omega_{c} \pm \sqrt{2}J$. The crossings are exact, because Eq.~(\ref{eq4}) holds for each supermode separately. The bundle therefore supports four independent lattices of microresonators. Neighboring microresonators belong to different supermodes and are a distance $\Lambda$ apart. The eigenmodes of each microresonator have equal intensity in the four fibers.

Figure~\ref{fig3} shows the eigenfrequencies and the axial intensity distributions of the modes for the characteristic $J/2\pi = 20$ GHz (for contacting 62.5~$\mu$m radius silica fibers~\cite{r34}) and $m = 353$. The depth $4J/2\pi = 80$ GHz corresponds to an effective radius variation of about 26 nm. For $\Lambda = 25$~$\mu$m shown in Fig.~\ref{fig3}(a), when the potential wells of each supermode have the period $4\Lambda = 100$~$\mu$m, each well has three eigenfrequencies. Tunneling between neighboring wells splits each of them into a band of closely spaced eigenfrequencies, with the widths 0.006, 0.20 and 2.2 GHz, and the lowest mode occupies 23~$\mu$m. This width is close to $\Lambda$, so neighboring microresonators overlap. They are not coupled, because they belong to different supermodes, which holds for identical fibers only (see below). For $\Lambda = 50$~$\mu$m shown in Fig.~\ref{fig3}(b), each potential well has seven eigenfrequencies, and the four lowest bands are narrower than 0.01 GHz. A band narrower than the resonance linewidth corresponds to isolated microresonators. A wider band corresponds to a chain of coupled microresonators with coinciding eigenfrequencies. For a linewidth of 0.1 GHz the microresonators of the lowest band of each lattice are isolated from one another for $\Lambda \gtrsim 19$~$\mu$m (SM, S7).

The spacing $\Lambda$ can be changed by the relative angle of the end holders or their axial translation (see \textit{Possible realizations} below). The window $60^{\circ} \leq \theta \leq 120^{\circ}$ allows at most $2m/3 \approx 235$ microresonators. A bundle sheared over the whole window has the length $L = 2m\Lambda/3$, which is 5.9 mm for $\Lambda = 25$~$\mu$m and 11.8 mm for $\Lambda = 50$~$\mu$m. Conversely, a bundle of length $L$ has the smallest spacing $\Lambda_{\min} = 3L/(2m)$ (Fig.~\ref{fig2}). The spacing of the lowest eigenfrequencies, $2m\sigma\sqrt{DJ}$, grows linearly with the gradient $\sigma$, and a change $\delta\theta$ of the end angles shifts the lattice by $\delta\theta/\sigma$ (SM, S7). A transverse microfiber in contact with one fiber excites the four supermodes equally, so the spectrogram measured along the bundle shows resonances that repeat with the period $\Lambda$ (SM, S8).

\section*{Unequal fibers}

If the cutoff frequency of fiber $j$ is shifted from $\omega_{c}$ by $\Delta\omega_{c,j}$, the supermodes are coupled, and the crossings of their cutoff frequencies at $\omega_{c} \pm \sqrt{2}J$ and at $\omega_{c}$ become avoided crossings. The smallest separations of the two cutoff frequencies at these avoided crossings (the gaps) are constants determined by the shifts of the cutoff frequencies of the fibers (SM, S5),
\begin{equation}\label{eq6}
\begin{aligned}
\delta_{\omega_{c}\pm\sqrt{2}J}&=\frac{1}{2}\sqrt{\left(\Delta\omega_{c,0}-\Delta\omega_{c,2}\right)^{2}+\left(\Delta\omega_{c,1}-\Delta\omega_{c,3}\right)^{2}},\\
\delta_{\omega_{c}}&=\frac{1}{2}\left|\Delta\omega_{c,0}-\Delta\omega_{c,1}+\Delta\omega_{c,2}-\Delta\omega_{c,3}\right|.
\end{aligned}
\end{equation}

A shift of the bundle by $\Lambda$ changes the phase $\phi(z)$ in Eq.~(\ref{eq2}) by $\pi/2$, which is compensated by the substitution $\psi_{j}(z) \rightarrow e^{- i\pi j/2}\psi_{j}(z)$ for any cutoff frequencies of the fibers (SM, Eq. (S19)). Each of the separated branches therefore has the period $\Lambda$ (Fig.~\ref{fig4}), in agreement with Eq.~(\ref{eq7}) below, according to which the local cutoff frequencies depend on the coupling phases only through $\Phi$. The four interleaved lattices of depth $4J$ survive for a mismatch much smaller than $J$ (for the required matching of the fiber radii, see \textit{Possible realizations} below). For a mismatch comparable with $J$ the lattice is transformed, and we solve Eq.~(\ref{eq2}) directly. With one fiber detuned by $\Delta\omega_{c,0}/2\pi = 20$ GHz the four branches have the modulation depths 8.0, 20.0, 18.9 and 6.9 GHz. These potential wells are shallower and four times narrower than for identical fibers, and separate microresonators appear only at larger spacings. Resonances confined to individual potential wells are absent at $\Lambda=25$ and 50~$\mu$m, and appear in the two deepest branches (20.0 and 18.9 GHz) at $\Lambda = 125$~$\mu$m and in all four branches at $\Lambda = 250$~$\mu$m (Fig.~\ref{fig4} and SM, S9). At $\Lambda = 250$~$\mu$m the lowest branch supports three bands with one state per potential well, with the widths 0.007, 0.10 and 0.60 GHz.

\begin{figure}[t]
\centering\includegraphics[width=\columnwidth]{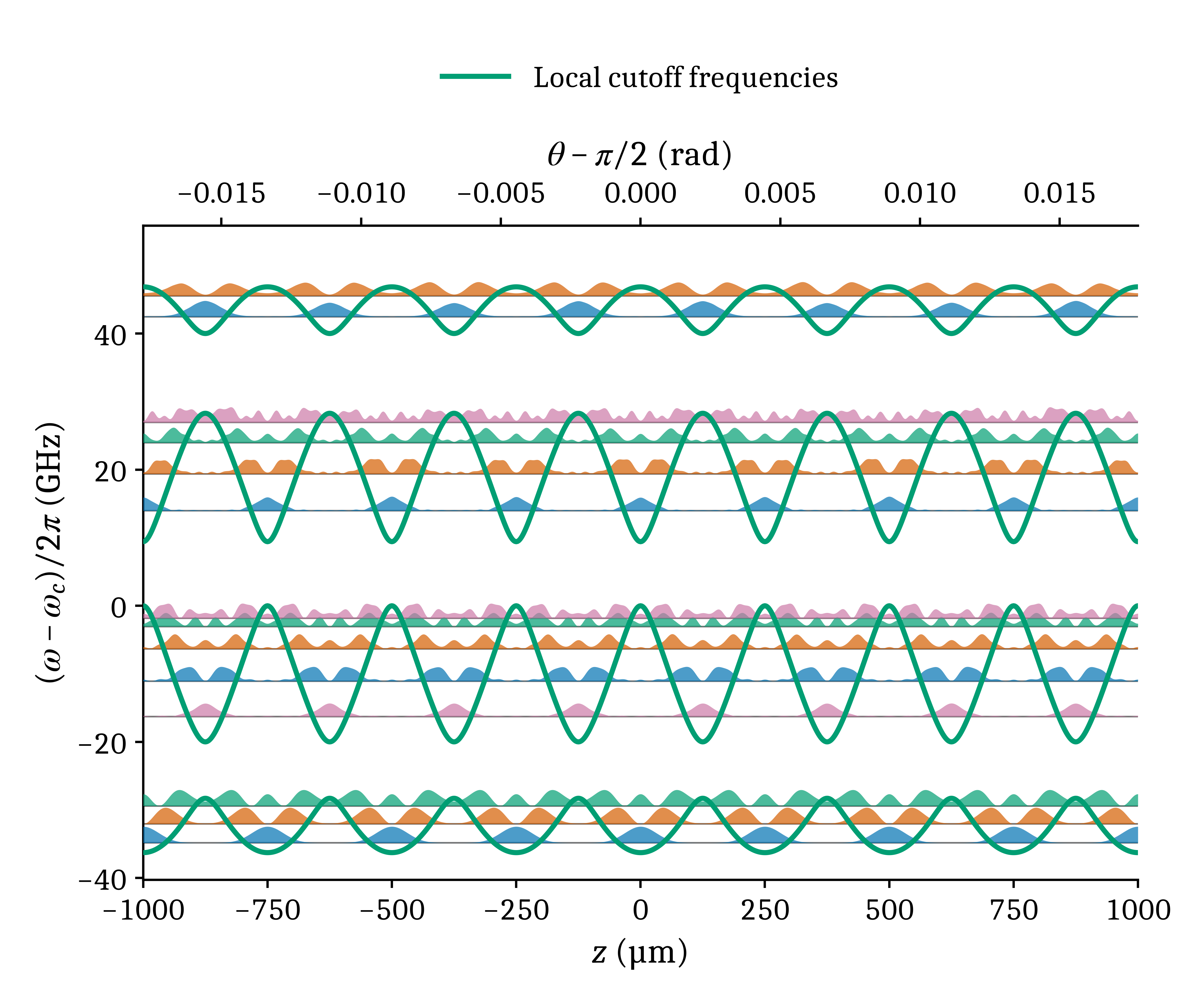}
\caption{\label{fig4}Localized states of a bundle with one fiber detuned by $\Delta\omega_{c,0}/2\pi=+20$ GHz for 40 periods of $\Lambda = 250$~$\mu$m, $J/2\pi=20$ GHz, with fully absorbing ends; the eight central periods are shown. Green: the four local cutoff frequencies, which have the period $\Lambda$. Horizontal lines: frequencies of the bands of localized states of Eq.~(\ref{eq2}). Shaded: axial intensity summed over the states of each band and over the four fibers. Upper axis: rhombus angle relative to the square.}
\end{figure}

The difference between the cutoff frequencies of the fibers can also be introduced and tuned on purpose, by stretching a fiber: a relative elongation of about $10^{- 3}$ shifts its cutoff frequency by about 100 GHz~\cite{r41}, so that the detuning $\Delta\omega_{c,0}$ of the stretched fiber covers the whole interval from $\Delta\omega_{c,0} \ll J$ to $\Delta\omega_{c,0} \gg J$. The microresonators are always $\Lambda$ apart, and the detuning changes the coupling between neighboring microresonators, which is zero for identical fibers. For one fiber detuned by $\Delta\omega_{c,0} \ll J$ their common eigenfrequency becomes a band, whose width is $0.48\Delta\omega_{c,0}$ at $\Lambda = 25$~$\mu$m and $0.21\Delta\omega_{c,0}$ at $\Lambda = 50$~$\mu$m (SM, S9).

\section*{Bundles of $N$ fibers}

We now generalize the results obtained above for four fibers to a bundle of $N \geq 3$ fibers whose contacts form a closed coupling loop (Fig.~\ref{fig1}(b)). We consider fibers with different cutoff frequencies $\omega_{c,j}$, the same coupling magnitude $J$ and arbitrary coupling phases $\varphi_{j}$. In this case, the local cutoff frequencies at a given $z$ are the eigenvalues $\Omega$ of the $N \times N$ matrix $\mathbf{H}$ with the elements $H_{jj} = \omega_{c,j}$ and $H_{j,j + 1} = H_{j + 1,j}^{*} = Je^{i\varphi_{j}}$ ($j$ mod $N$), all other elements being zero. We prove that (SM, S2)
\begin{equation}\label{eq7}
\begin{aligned}
\det\left(\Omega-\mathbf{H}\right)&=P_{N}\left(\Omega;\{\omega_{c,j}\},J\right)-2J^{N}\cos\Phi,\\
\Phi&=\sum_{j}\varphi_{j},
\end{aligned}
\end{equation}
where the polynomial $P_{N}$ does not contain the phases. The local cutoff frequencies therefore depend on the phases only through $\Phi$, for any cutoff frequencies of the fibers, in agreement with the period $\Lambda$ found above after Eq.~(\ref{eq6}) for four unequal fibers. For an open chain the last term is absent. If the azimuthal angle in each fiber is measured from the same fixed transverse axis, $\varphi_{j} = \left( m_{j + 1} - m_{j} \right)\vartheta_{j,j + 1}$, where $\vartheta_{j,j + 1}$ is the angle between the line of centers of fibers $j$ and $j + 1$ and this axis. A twist of the bundle, which rotates its cross-section rigidly by the angle $\gamma(z)$, changes the phase of each coupling by $(m_{j + 1} - m_{j})\gamma$, which is $\mp 2m\gamma$ for alternating azimuthal numbers. It does not change $\Phi$ for any $N$ and any twist rate, because the sum of $m_{j + 1} - m_{j}$ around a closed loop is zero (SM, S4). Hence $\Phi$ depends only on the cross-sectional configuration of the bundle. This is in contrast with twisted multicore fibers~\cite{r27}, where the twist changes the loop phase of the core modes. For an even number of fibers with alternating circulations $\Phi \equiv 2m\sum_{j\,\mathrm{even}}\theta_{j}$, where $\theta_{j}$ is the interior angle at vertex $j$ of the polygon formed by the fiber centers. For the rhombus, $\theta_{0} = \theta_{2} = \theta$, which gives Eq.~(\ref{eq3}), and the phases $\varphi_{j}$ reduce, up to the constants $\eta_{j}$, to $\phi(z)$ of Eq.~(\ref{eq2}). The WGMs with the opposite circulations, $m_{j} = - ( - 1)^{j}m$, form a second loop, which is not coupled to the first one and has the loop phase $- \Phi$. It follows from Eq.~(\ref{eq7}) that its local cutoff frequencies are the same, so every state of the bundle is doubly degenerate. For an odd number of fibers, the circulations cannot alternate around the loop: after one turn light returns to the initial fiber with the opposite circulation. The two loops then merge into a single loop of $2N$ modes, which closes after two turns around the bundle. On the second turn each contact is passed with the two circulations interchanged and contributes the opposite phase, so $\Phi$ vanishes for any configuration of the bundle (SM, S4). Four fibers are therefore the smallest bundle with a shape-dependent geometric phase. Eq.~(\ref{eq2}) with $j$ mod $N$ holds for a bundle of an even number $N$ of fibers when all coupling phases change at the same rate, and in the adiabatic approximation otherwise (SM, S3). In such a bundle, neighboring microresonators are separated by the distance over which $\Phi$ changes by $2\pi$.

\section*{Possible realizations}

The configuration shown in Fig.~\ref{fig2} can be realized, e.g., by holding the four fibers, each under the same axial tension, in two end holders with elastic V-groove fingers that set the cross-section and the direction of each fiber. The coupling magnitude $J$ changes by about 0.4\% per nanometer of the gap between the fibers, so the fibers must be in direct contact along the whole bundle (SM, S6). The holders bring the fibers into contact at the bundle ends. Between the ends, the contact is maintained by the attraction between the silica surfaces, which was observed to keep two fibers in direct contact over millimeter lengths~\cite{r39}, complemented by the axial tension of each fiber, which presses adjacent fibers against each other with a negligible flattening of the contacts~\cite{r43,r44,r45} (SM, S6). The attraction, which makes fibers in contact adhere~\cite{r46}, is dominated by van der Waals forces at nanometer separations~\cite{r39,r47,r48}, with a longer-range electrostatic contribution~\cite{r47,r48} (SM, S6). The relative tilt of adjacent fibers, $r_{0}\sigma$, is constant along the bundle and equals 11 mrad for $\Lambda = 25$~$\mu$m and 5.6 mrad for $\Lambda = 50$~$\mu$m [Figs.~\ref{fig3}(a) and \ref{fig3}(b)]. In contrast to the tilt of two intersecting fibers~\cite{r39}, it rotates the line of centers of adjacent fibers and does not separate them. A change of the fiber radius by 1 nm shifts the cutoff frequency by about 3 GHz, so the lattices of identical fibers need radii matched to much better than 6 nm (see Eq.~(\ref{eq6})). This is possible since segments of the same parent fiber can be matched to 0.1 GHz~\cite{r49} and the residual mismatch can be trimmed by stretching~\cite{r41}. A bundle with a larger even $N$ can be held in the same way, with the shape of its cross-section set at each end by nanopositioners.

\section*{Conclusion}

We have shown that the geometric phase of WGMs in a bundle of side-coupled optical fibers creates a lattice of microresonators whose period is much shorter than the length over which the bundle changes. The loop phase depends on the cross-sectional configuration of the bundle and is invariant under rigid twist. In four identical fibers sheared from a rhombus through a square to the rotated rhombus, it gives four interleaved lattices of depth $4J$ and period $4\Lambda$, in which neighboring microresonators are a distance $\Lambda = \pi/(2m|\sigma|)$ apart. The end holders can tune the microresonator period $4\Lambda$ down to about 100~$\mu$m. A difference between the cutoff frequencies of the fibers, which can be set by stretching, couples the neighboring microresonators and, when it is comparable with $J$, replaces the four lattices by one lattice of period $\Lambda$. The lattice requires a closed coupling loop with an \emph{even number of fibers}, and every state is doubly degenerate, because the WGMs of the opposite circulation form a second, uncoupled loop. For an odd number of fibers, the WGMs of the two circulations form a single loop, in which the coupling phases cancel, so the loop phase vanishes for any configuration of the bundle. The same mechanism applies to more general closed loops of an even number of coupled WGM resonators illustrated in Fig.~\ref{fig1}(b). Our model includes only the nearest-neighbor coupling of one WGM in each fiber, with given azimuthal and radial numbers and polarization. It neglects other WGMs, scattering at the fiber surfaces and the flattening of the contacts (SM, S10), the consideration of which is left for future work.

\section*{Acknowledgments}

This work was supported by the Engineering and Physical Sciences Research Council (EPSRC) (Grant Nos. EP/X03772X/1 and EP/W002868/1).

\section*{Data availability}

The data that support the findings of this study are available from the corresponding author upon reasonable request.

\clearpage
\onecolumn
\setcounter{section}{0}\setcounter{equation}{0}\setcounter{figure}{0}\setcounter{table}{0}
\setcounter{secnumdepth}{3}
\renewcommand{\thesection}{S\arabic{section}}
\renewcommand{\theequation}{S\arabic{equation}}
\renewcommand{\thefigure}{S\arabic{figure}}
\renewcommand{\thetable}{S\arabic{table}}
\renewcommand{\theHsection}{S\arabic{section}}
\renewcommand{\theHequation}{S\arabic{equation}}
\renewcommand{\theHfigure}{S\arabic{figure}}
\renewcommand{\theHtable}{S\arabic{table}}

\begin{center}
\textbf{\large Supplemental Material:\\ Tunable Microresonator Arrays Induced by Geometric Phase}
\end{center}
\vspace{2ex}

Reference numbers refer to the reference list of the main text, and equations without the prefix S are those of the main text. Sections~\ref{secS1} to \ref{secS6} derive the model and its general properties, and Sections~\ref{secS7} to \ref{secS10} give the numerical results and the details of the lattices that are summarized in the main text. Symbols: $\psi_{j}(z)$ is the axial envelope of the WGM in fiber $j$; $\chi_{\ell }(z)$ is the amplitude of the supermode $\ell$ (Section~\ref{secS3}); $\vartheta_{jk}$ is the direction of the line of centers from fiber $j$ to fiber $k$; $\eta_{j}(z)$ is the phase of a redefinition of the modes; $\Omega$ is the eigenvalue variable of the cutoff problem; bold symbols denote vectors and matrices. All frequencies in the equations are angular frequencies, and the numerical values given in GHz are these quantities divided by $2\pi$.

\section{Coupled wave equations for a bundle of $N$ fibers}\label{secS1}

We consider $N \geq 3$ circular fibers of equal radius whose centers $\mathbf{R}_{j}$ form a closed polygon in each transverse plane. The fibers are numbered $j = 0,\ldots,N - 1$ (mod $N$) counterclockwise, and each fiber is coupled to its two neighbors. We call the closed chain of these couplings the \textit{coupling loop}. The direction of the line of centers from fiber $j$ to fiber $k$ is given by the angle $\vartheta_{jk}$ between the vector $\mathbf{R}_{k} - \mathbf{R}_{j}$ and the $x$ axis, measured counterclockwise, so that $\mathbf{R}_{k} - \mathbf{R}_{j} = \left| \mathbf{R}_{k} - \mathbf{R}_{j} \right|\left( \cos\vartheta_{jk},\ \sin\vartheta_{jk} \right)$ and $\vartheta_{kj} = \vartheta_{jk} + \pi$. Each fiber carries one family of WGMs with the azimuthal number $m_{j} = \pm m$ and the cutoff frequency $\omega_{c,j}$, which is $\omega_{c}$ for identical fibers and $\omega_{c} + \Delta\omega_{c,j}$ for unequal fibers.

Graf's addition theorem [42] expands the exterior field of the mode $m_{k}$ of fiber $k$ about fiber $j$. The coupling of the modes $m_{j}$ and $m_{k}$ then contains the Hankel function of the order $m_{k} - m_{j}$ and the phase factor $e^{i\left( m_{k} - m_{j} \right)\left( \vartheta_{jk} + \pi \right)}$. The order is $0$ for equal and $2m$ for opposite circulations, and the factor $\pi$ drops out because $m_{k} - m_{j}$ is even. For high-$m$ modes the opposite-circulation coupling dominates by many orders of magnitude, and we retain only it. The envelopes $\psi_{j}(z)$ then obey the equation [34,35]
\begin{equation}\label{eqS1}
\begin{gathered}
D\frac{d^{2}\psi_{j}}{dz^{2}} + \left( \omega - \omega_{c,j} \right)\psi_{j} - \sum_{k = j \pm 1}h_{jk}\psi_{k} = 0,\qquad  h_{jk} = J_{jk}\, e^{i\left( m_{k} - m_{j} \right)\vartheta_{jk}}, \\
D = c^{2}/2n_{0}^{2}\omega_{c},\qquad h_{kj} = h_{jk}^{*}.
\end{gathered}
\end{equation}

The magnitude $J_{jk}$ decreases exponentially with the gap between the fibers [34]. We assume the same magnitude $J$ for all links, which holds when adjacent fibers are in direct contact along the whole bundle, and write $h_{j,j + 1} = Je^{i\varphi_{j}}$ with
\begin{equation}\label{eqS2}
\varphi_{j} = \left( m_{j + 1} - m_{j} \right)\,\vartheta_{j,j + 1}.
\end{equation}

For two fibers tilted by $\alpha$ the line of centers rotates by $\alpha z/2r_{0}$. With $\left| m_{2} - m_{1} \right| = 2m$, Eq. (\ref{eqS2}) gives the phase $\beta_{0}z$, $\beta_{0} = 2\pi\alpha n_{0}/\lambda \approx m\alpha/r_{0}$, of the coupling coefficient in Eq. (7) of Ref. [34].

A redefinition of the phases of the modes, $\psi_{j}(z) \rightarrow e^{i\eta_{j}}\psi_{j}(z)$ with constant $\eta_{j}$ changes $\varphi_{j} \rightarrow \varphi_{j} + \eta_{j + 1} - \eta_{j}$ and leaves the sum $\Phi = \sum_{j}\varphi_{j}$ unchanged. At a fixed axial position, for an open chain every $\varphi_{j}$ can be removed in this way, and for a closed loop only $\Phi$ (mod $2\pi$) remains. We call $\Phi$ the \textit{loop phase}.

\textit{Two fibers: the effect of the coupling phase.} For an open chain the phases of the couplings do not form a loop phase, but they are not eliminated either when they change along the fibers. The simplest example is two fibers with the same cutoff frequency $\omega_{c}$ and the coupling $Je^{i\beta_{0}z}$ of Ref. [34]. Equation (\ref{eqS1}) then reads
\begin{equation*}
D\frac{d^{2}\psi_{1}}{dz^{2}} + \left( \omega - \omega_{c} \right)\psi_{1} - Je^{i\beta_{0}z}\psi_{2} = 0,\qquad D\frac{d^{2}\psi_{2}}{dz^{2}} + \left( \omega - \omega_{c} \right)\psi_{2} - Je^{- i\beta_{0}z}\psi_{1} = 0.
\end{equation*}

The substitution $\psi_{1} = e^{i\beta_{0}z/2}\chi_{1}$, $\psi_{2} = e^{- i\beta_{0}z/2}\chi_{2}$ removes the phase from the coupling and transfers it to the derivatives,
\begin{equation*}
D\left( \frac{d}{dz} \pm \frac{i\beta_{0}}{2} \right)^{2}\chi_{1,2} + \left( \omega - \omega_{c} \right)\chi_{1,2} - J\chi_{2,1} = 0,
\end{equation*}
where the upper sign refers to $\chi_{1}$ and the lower sign to $\chi_{2}$. The phase is therefore equivalent to a displacement of the axial wavenumbers of the two WGMs by $\beta_{0}$ relative to each other. For a constant $J$ these equations do not depend on $z$, and their solutions $\chi_{1,2} \propto e^{ikz}$ have the frequencies
\begin{equation*}
\omega_{\pm}(k) = \omega_{c} + D\left( k^{2} + \frac{\beta_{0}^{2}}{4} \right) \pm \sqrt{J^{2} + D^{2}\beta_{0}^{2}k^{2}}.
\end{equation*}

For $\beta_{0} \ll \sqrt{\frac{J}{D}}$ the expansion in $k$ gives
\begin{equation*}
\omega_{\pm}(k) \approx \omega_{c} \pm J + \frac{D\beta_{0}^{2}}{4} + D\left( 1 \pm \frac{D\beta_{0}^{2}}{2J} \right)k^{2},
\end{equation*}
so the phase only shifts both frequencies by $\frac{D\beta_{0}^{2}}{4}$ and changes the coefficient $D$ by the small relative amount $\frac{D\beta_{0}^{2}}{2J}$. In this limit a microresonator formed by a variation of $J$ along the fibers is described by the potential $\omega_{c} \mp J(z)$, as in the adiabatic approximation. For larger $\beta_{0}$ the phase takes part in forming the eigenmodes. In particular, for $D\beta_{0}^{2} > 2J$ the lower branch has two minima, at $k = \pm \left( \frac{\beta_{0}}{2} \right)\sqrt{1 - \frac{4J^{2}}{D^{2}\beta_{0}^{4}}}$, where $\omega_{-} = \omega_{c} - \frac{J^{2}}{D\beta_{0}^{2}}$, and the eigenmodes of a microresonator are then composed of waves with these wavenumbers. For $J/2\pi = 20$ GHz, $\sqrt{\frac{J}{D}} \approx 0.084$ $\mu\mathrm{m}^{-1}$. The tilt of 11 mrad between adjacent fibers at $\Lambda = 25$~$\mu$m corresponds to $\beta_{0} \approx 0.063$ $\mu\mathrm{m}^{-1}$, for which $\frac{D\beta_{0}^{2}}{4} \approx 2\pi \times 2.8$ GHz and $\frac{D\beta_{0}^{2}}{2J} \approx 0.28$. In any case a phase that grows linearly along two fibers does not make the structure periodic, because it enters the transformed equations only through the constant $\beta_{0}$. In the bundle of four identical fibers the phase enters differently: there the separation into Eq. (\ref{eqS6}) is exact, and the phase appears in the potential of each supermode.

\section{Invariance of the loop phase}\label{secS2}

At a fixed $z$ the local cutoff frequencies are the eigenvalues of the $N \times N$ Hermitian matrix $\mathbf{H}$ with $H_{jj} = \omega_{c,j}$, $H_{j,j + 1} = Je^{i\varphi_{j}}$, $H_{j + 1,j} = Je^{- i\varphi_{j}}$ and all other elements zero. We prove that for $N \geq 3$, arbitrary cutoff frequencies $\omega_{c,j}$ and phases $\varphi_{j}$, and the same $J$ on all links,
\begin{equation}\label{eqS3}
\det\left( \Omega\mathbf{I} - \mathbf{H} \right) = \sum_{M}\left( - J^{2} \right)^{|M|}\prod_{j \notin M}\left( \Omega - \omega_{c,j} \right)\mspace{6mu} - \mspace{6mu} 2J^{N}\cos\Phi,\qquad \Phi = \sum_{j = 0}^{N - 1}\varphi_{j}.
\end{equation}

Here $M$ runs over all sets of non-overlapping links of the loop (including the empty set), $|M|$ is the number of links in $M$, and the product runs over the fibers not covered by $M$. The first term is the polynomial $P_{N}\left( \Omega;\{\omega_{c,j}\},J \right)$ of the main text. It does not contain the phases.

To prove Eq. (\ref{eqS3}), we put $\mathbf{X} = \Omega\mathbf{I} - \mathbf{H}$, where $\mathbf{I}$ is the unit matrix, so that $X_{jj} = \Omega - \omega_{c,j}$, $X_{j,j + 1} = - Je^{i\varphi_{j}}$ and $X_{j + 1,j} = - Je^{- i\varphi_{j}}$. In the expansion $\det\mathbf{X} = \sum_{\pi}\mathrm{sgn}(\pi)\prod_{j}X_{j\pi(j)}$ a permutation contributes only if $\pi(j) \in \{ j,j \pm 1\}$ for all $j$. We split $\pi$ into cycles. A cycle of length 1 is a fixed fiber. A cycle of length 2 exchanges two neighbors. A longer cycle moves by steps of $\pm 1$ along the loop without revisiting a fiber, so it can only close by going once around the whole loop, which has length $N$ and two senses. The permutations that contribute are therefore the products of fixed fibers and exchanges of neighbors, and the two cyclic permutations $j \rightarrow j \pm 1$.

A product of fixed fibers and exchanges corresponds to a set $M$ of exchanged links. Its weight is $\prod_{j \notin M}\left( \Omega - \omega_{c,j} \right)\prod_{(j,j + 1) \in M}X_{j,j + 1}X_{j + 1,j}$ and its sign is $( - 1)^{|M|}$. Since $X_{j,j + 1}X_{j + 1,j} = J^{2}e^{i\varphi_{j}}e^{- i\varphi_{j}} = J^{2}$, the phases cancel and we obtain the first term of (\ref{eqS3}). For the cycle $j \rightarrow j + 1$ the product is $\prod_{j}X_{j,j + 1} = ( - J)^{N}e^{i\Phi}$ and the sign of an $N$-cycle is $( - 1)^{N - 1}$, so the contribution is $- J^{N}e^{i\Phi}$. The opposite cycle gives $- J^{N}e^{- i\Phi}$, and the sum is $- 2J^{N}\cos\Phi$.

The phases enter the local spectrum only through $\cos\Phi$. Any redistribution of the $\varphi_{j}$ over the links that keeps $\Phi$ leaves the local cutoff frequencies unchanged, whether or not it is a redefinition of the phases of the modes.

Equation (\ref{eqS3}) has the following consequences.

\begin{enumerate}
\item
  For equal cutoff frequencies $\omega_{c,j} = \omega_{c}$ the roots of (\ref{eqS3}) are $\Omega_{\ell } = \omega_{c} + 2J\cos\left[ \left( \Phi + 2\pi\ell  \right)/N \right]$, $\ell =0,\ldots,N - 1$. A change of $\Phi$ by $2\pi$ permutes the branches, and each branch has the period $2\pi N$ in $\Phi$.
\item
  For a nondegenerate local cutoff frequency $\Omega_{\nu}$, i.e., away from the crossings of the branches, differentiating $P_{N}(\Omega) - 2J^{N}\cos\Phi = 0$ gives $\partial\Omega_{\nu}/\partial\cos \Phi = 2J^{N}/\left( dP_{N}/d\Omega \right)_{\Omega = \Omega_{\nu}}$. At the lowest root $dP_{N}/d\Omega$ has the sign $( - 1)^{N - 1}$, so for even $N$ the lowest cutoff frequency decreases monotonically with $\cos\Phi$, and as $\Phi(z)$ changes it forms a periodic sequence of potential wells, also for different $\omega_{c,j}$.
\item
  For different magnitudes $J_{j}$ the same proof gives $J^{N} \rightarrow \prod_{j}J_{j}$ in the last term and $J^{2} \rightarrow J_{j}^{2}$ in the matching terms, so the invariance of $\Phi$ holds as well.
\end{enumerate}

\section{When does the problem separate into independent equations?}\label{secS3}

Equation (\ref{eqS3}) is exact at a fixed $z$. In terms of the column vector $\bm{\psi}(z) = \left( \psi_{0},\ \psi_{1},\ \ldots,\ \psi_{N - 1} \right)^{T}$ of the envelopes in the $N$ fibers and the matrix $\mathbf{H}(z)$ of Section~\ref{secS2}, the axial coupled wave equations (\ref{eqS1}) can be written as
\begin{equation}\label{eqS4}
D\frac{d^{2}\bm{\psi}}{dz^{2}} + \left[ \omega\mathbf{I} - \mathbf{H}(z) \right]\bm{\psi} = 0.
\end{equation}

The question is when its solutions are determined by the local roots of (\ref{eqS3}) alone.

\textit{General criterion.} Let $\mathbf{e}_{\nu}(z)$ be the normalized eigenvectors of $\mathbf{H}(z)$ and $\Omega_{\nu}(z)$ the corresponding eigenvalues, which are the local cutoff frequencies, and write $\bm{\psi} = \sum_{\nu}f_{\nu}(z)\mathbf{e}_{\nu}(z)$. Then $f_{\nu}$ obeys $D\left( d/dz + A_{\nu\nu} \right)^{2}f_{\nu} + \left( \omega - \Omega_{\nu} \right)f_{\nu} = 0$ plus coupling to the other states through $A_{\nu\mu} = \langle\mathbf{e}_{\nu}|d\mathbf{e}_{\mu}/dz\rangle$. The local cutoff frequencies $\Omega_{\nu}(z)$ are exact potentials if and only if the eigenvectors do not depend on $z$, so that $A_{\nu\mu} = 0$. The adiabatic approximation neglects $A_{\nu\mu}$ with $\nu \neq \mu$.

\textit{Equal cutoff frequencies, equal couplings, equal rates.} Let $\omega_{c,j} = \omega_{c}$, $J_{j} = J$, and let the phases change at the same rate, $d\varphi_{j}/dz = d\phi/dz$ for all $j$. Constant differences between the $\varphi_{j}$ can be removed by a constant redefinition of the phases, so we may take $\varphi_{j}(z) = \phi(z)$, with $\Phi = N\phi$. Then $\mathbf{H} = \omega_{c}\mathbf{I} + J\left( e^{i\phi}\bm{\Pi} + e^{- i\phi}\bm{\Pi}^{\dagger} \right)$ with the cyclic shift $\bm{\Pi}$, and its eigenvectors are the Fourier vectors $N^{- 1/2}e^{2\pi i\ell j/N}$, which do not depend on $z$. We therefore write
\begin{equation}\label{eqS5}
\psi_{j}(z) = N^{- 1/2}\sum_{\ell =0}^{N - 1}e^{2\pi i\ell j/N}\chi_{\ell }(z),
\end{equation}
where $\chi_{\ell }(z)$ is the amplitude of the supermode $\ell$, in which all fibers carry the same intensity and neighboring fibers differ in phase by $2\pi\ell /N$. It is not the amplitude in a fiber. The $N$ equations decouple exactly:
\begin{equation}\label{eqS6}
D\frac{d^{2}\chi_{\ell }}{dz^{2}} + \left[ \omega - \omega_{c} - 2J\cos\left( \phi(z) + 2\pi\ell /N \right) \right]\chi_{\ell } = 0.
\end{equation}

For $N = 4$ this is Eq. (4) of the main text. The decoupling does not require $d\phi/dz$ to be small, so the potentials are exact for fast as well as slow changes, within the nearest-neighbor model. For a linear change, $\phi = Kz$ with $K = m\sigma$, and $x = \left( Kz + 2\pi\ell /N \right)/2$ turns (\ref{eqS6}) into the Mathieu equation $d^{2}\chi/dx^{2} + \left( a_{M} - 2q_{M}\cos 2x \right)\chi = 0$ with $a_{M} = 4\left( \omega - \omega_{c} \right)/DK^{2}$ and $q_{M} = 4J/DK^{2}$. The potential well depth is $4J$ for every $N$, the period of the cosine in $z$ is $2\pi/K$, and the $N$ supermodes have identical lattices displaced by $2\pi/(NK)$ along $z$, which is the distance $\Lambda$ of the main text for $N = 4$.

\textit{Which deformations have equal rates.} For $N$ even and alternating circulations, $m_{j} = ( - 1)^{j}m$, Eq. (\ref{eqS2}) gives $d\varphi_{j}/dz = - 2m( - 1)^{j}d\vartheta_{j,j + 1}/dz$. The rates are equal if $( - 1)^{j}d\vartheta_{j,j + 1}/dz$ is the same for all $j$. That is, the directions of the even links rotate in one sense and those of the odd links in the opposite sense at the same angular speed. For the loop to remain closed for all values of the rotation, the unit vectors along the even links must sum to zero, and so must those along the odd links.

\textit{Coupling phases of the rhombus.} For $N = 4$ these conditions are satisfied by the rhombus. The sides $0 \rightarrow 1$ and $2 \rightarrow 3$ are antiparallel and the sides $1 \rightarrow 2$ and $3 \rightarrow 0$ are antiparallel. With the centers $(a,0),(0,b),( - a,0),(0, - b)$ and $a = 2r_{0}\cos(\theta/2)$, $b = 2r_{0}\sin(\theta/2)$, where $r_{0}$ is the fiber radius, the four directions are $\pi - \theta/2$, $\pi + \theta/2$, $- \theta/2$ and $\theta/2$, and Eq. (\ref{eqS2}) gives $\varphi_{j} = m\theta$ (mod $2\pi$) on every link. All rates are equal to $m\, d\theta/dz$ for any $\theta(z)$. At the square, $\theta = \pi/2$, the common phase is $m\pi/2$, which a constant redefinition of the phases removes because $4 \cdot m\pi/2 = 2\pi m$ is a multiple of $2\pi$. After this redefinition $\phi = m(\theta - \pi/2)$, as in Eq. (2) of the main text, and the loop phase is $\Phi = 4\phi = 4m(\theta - \pi/2)$ (mod $2\pi$), which is Eq. (3) of the main text. A hexagon with even and odd links rotating in opposite senses (for example from the regular hexagon, with three directions $0^{\circ},120^{\circ},240^{\circ}$ for the even links and $60^{\circ},180^{\circ},300^{\circ}$ for the odd links) is a second example: the unit vectors along its even links sum to zero for any rotation angle, and so do those along its odd links.

\textit{Unequal rates.} If the rates $d\varphi_{j}/dz$ differ but $\Phi(z)$ is the same, the local roots are still $\omega_{c} + 2J\cos\left[ \left( \Phi + 2\pi\ell  \right)/N \right]$, as follows from Eq. (\ref{eqS3}), but the eigenvectors depend on $z$. A $z$-dependent redefinition of the phases, $\bm{\psi} = \mathbf{U}\bm{\chi}$, $\mathbf{U} = \mathrm{diag}\left( e^{i\eta_{j}(z)} \right)$, can make the phases equal. It then changes the kinetic term to $D\left( d/dz + i\mathcal{A} \right)^{2}$ with $\mathcal{A} = \mathrm{diag}\left( d\eta_{j}/dz \right)$, which couples the supermodes. The potentials of (\ref{eqS6}) are then valid only in the adiabatic approximation, that is, for $\left| d\varphi_{j}/dz - d\phi/dz \right| \ll \sqrt{J/D}$, where $z_{0} = \sqrt{D/J}$ is the length at which the kinetic energy $D/z_{0}^{2}$ equals $J$. In this case the individual phase profiles, and not only $\Phi$, determine the axial dynamics.

\textit{Unequal cutoff frequencies.} For $\omega_{c,j} = \omega_{c} + \Delta\omega_{c,j}$ the Fourier vectors are no longer eigenvectors. In the basis (\ref{eqS5}) the equations become
\begin{equation}\label{eqS7}
\begin{gathered}
D\frac{d^{2}\chi_{\ell }}{dz^{2}} + \left[ \omega - \omega_{c} - 2J\cos\left( \phi + 2\pi\ell /N \right) \right]\chi_{\ell } - \sum_{\ell' \neq \ell }{\widehat{\Delta}}_{\ell - \ell'}\chi_{\ell'} = 0, \\
{\widehat{\Delta}}_{q} = \frac{1}{N}\sum_{j}\Delta\omega_{c,j}e^{- 2\pi iqj/N},
\end{gathered}
\end{equation}
and the supermodes are coupled by the Fourier components of the cutoff differences. This is the case treated in Section~\ref{secS5}.

\section{Rigid twist, shape dependence and odd bundles}\label{secS4}

\textit{Rigid rotation is a redefinition of the phases.} A rigid rotation of the bundle by $\gamma(z)$ adds $\gamma$ to every $\vartheta_{jk}$, and it follows from Eq. (\ref{eqS2}) that
\begin{equation}\label{eqS8}
\varphi_{j} \rightarrow \varphi_{j} + \left( m_{j + 1} - m_{j} \right)\gamma,\qquad \Phi \rightarrow \Phi + \gamma\sum_{j}\left( m_{j + 1} - m_{j} \right) = \Phi.
\end{equation}

The sum in (\ref{eqS8}) vanishes on a closed loop for any $N$ and any $m_{j}$, because each $m_{j}$ enters it twice, with opposite signs. The transformation is the redefinition of the phases $\psi_{j}(z) \rightarrow e^{- im_{j}\gamma}\psi_{j}(z)$, and it follows from Eq. (\ref{eqS3}) that the local cutoff frequencies do not change. In the rotating basis, $\bm{\psi} = \mathbf{Q}\bm{\chi}$ with $\mathbf{Q} = e^{- i\mathbf{P}\gamma(z)}$ and $\mathbf{P} = \mathrm{diag}\left( m_{j} \right)$, Eq. (\ref{eqS4}) has the kinetic term
\begin{equation}\label{eqS9}
D\left( \frac{d}{dz} - i\kappa\mathbf{P} \right)^{2},\qquad \kappa = \frac{d\gamma}{dz}.
\end{equation}

A twist therefore only shifts the axial momentum of the WGM in fiber $j$ by $m_{j}\kappa$.

\textit{Dependence on shape.} Summation by parts of $\Phi = \sum_{j}\left( m_{j + 1} - m_{j} \right)\vartheta_{j,j + 1}$ gives
\begin{equation}\label{eqS10}
\Phi = - \sum_{j}m_{j}\tau_{j}\quad(\mathrm{mod}\ 2\pi),\qquad \tau_{j} = \vartheta_{j,j + 1} - \vartheta_{j - 1,j},
\end{equation}
where $\tau_{j}$ is the turning angle of the polygon at fiber $j$. The ambiguity of $2\pi$ in $\tau_{j}$ does not matter because $m_{j}$ is an integer. For a convex polygon $\tau_{j} = \pi - \theta_{j}$ with the interior angle $\theta_{j}$. For an even $N$ with alternating circulations,
\begin{equation}\label{eqS11}
\Phi = m\sum_{j}( - 1)^{j}\theta_{j} \equiv 2m\sum_{j\,\mathrm{even}}\theta_{j}\quad(\mathrm{mod}\ 2\pi),
\end{equation}
where we used $\sum_{j}\theta_{j} = (N - 2)\pi$ and the integer $m$. For the rhombus, $\theta_{0} = \theta_{2} = \theta$, so that $\Phi \equiv 4m\theta\ (\mathrm{mod}\ 2\pi)$, which is Eq. (3) of the main text.

\textit{Odd bundles.} The sequence $m_{0} = m$, $m_{1} = - m$, $m_{2} = m,\ldots$ cannot close on an odd loop. The retained couplings connect the mode $(j,s)$, $s = \pm$, with the azimuthal number $sm$, to the mode $(j + 1, - s)$. These $2N$ modes form a single loop that closes only after two turns around the bundle. On the second turn every $m_{j}$ is replaced by $- m_{j}$ while the directions $\vartheta_{j,j + 1}$ are the same, so the contributions $\left( m_{j + 1} - m_{j} \right)\vartheta_{j,j + 1}$ of the two turns cancel, and the total phase of the loop of $2N$ modes vanishes for any shape. The cutoff frequencies are then $\omega_{c} + 2J\cos\left( \pi\ell /N \right)$ for equal fibers, independent of the shape and of the twist. The same-circulation coupling closes a loop of $N$ modes, but it is smaller than the retained coupling by many orders of magnitude and we neglect it. Four fibers form the smallest loop in which the shape produces a phase.

\textit{Example: three fibers.} We write out the smallest odd bundle explicitly. Three fibers 0, 1 and 2 with the same cutoff frequency $\omega_{c}$ touch one another at the contacts 0--1, 1--2 and 2--0, which have the coupling magnitudes $J_{1}$, $J_{2}$ and $J_{3}$. Each fiber carries two WGMs, with the azimuthal numbers $+ m$ and $- m$, so the bundle has six modes $(j,s)$. A contact couples only modes of opposite circulations. Starting from the mode $(0, + )$, light therefore passes the modes in the order
\begin{equation*}
(0, + )\  \rightarrow \ (1, - )\  \rightarrow \ (2, + )\  \rightarrow \ (0, - )\  \rightarrow \ (1, + )\  \rightarrow \ (2, - )\  \rightarrow \ (0, + ).
\end{equation*}

After three steps light returns to fiber 0, but in the mode $(0, - )$ and not in the initial mode $(0, + )$. The loop closes only after three more steps, in which the same three contacts are passed again. We number the modes $k = 1,\ldots,6$ in this order. It follows from Eq. (\ref{eqS2}) that the coupling phases on the first turn are
\begin{equation*}
\varphi_{1} = - 2m\vartheta_{01},\qquad \varphi_{2} = 2m\vartheta_{12},\qquad \varphi_{3} = - 2m\vartheta_{20}.
\end{equation*}

On the second turn the directions of the lines of centers are the same, and both azimuthal numbers at each contact have the opposite signs, so the phases are $- \varphi_{1}$, $- \varphi_{2}$ and $- \varphi_{3}$. The matrix $\mathbf{H}$ of Section~\ref{secS2} for the six modes is
\begin{equation*}
\mathbf{H} = \begin{pmatrix}
\omega_{c} & J_{1}e^{i\varphi_{1}} & 0 & 0 & 0 & J_{3}e^{i\varphi_{3}} \\
J_{1}e^{- i\varphi_{1}} & \omega_{c} & J_{2}e^{i\varphi_{2}} & 0 & 0 & 0 \\
0 & J_{2}e^{- i\varphi_{2}} & \omega_{c} & J_{3}e^{i\varphi_{3}} & 0 & 0 \\
0 & 0 & J_{3}e^{- i\varphi_{3}} & \omega_{c} & J_{1}e^{- i\varphi_{1}} & 0 \\
0 & 0 & 0 & J_{1}e^{i\varphi_{1}} & \omega_{c} & J_{2}e^{- i\varphi_{2}} \\
J_{3}e^{- i\varphi_{3}} & 0 & 0 & 0 & J_{2}e^{i\varphi_{2}} & \omega_{c}
\end{pmatrix}.
\end{equation*}

Each contact enters this matrix twice, with opposite phases: the contact 0--1 in the elements $H_{12}$ and $H_{45}$, the contact 1--2 in $H_{23}$ and $H_{56}$, and the contact 2--0 in $H_{34}$ and $H_{61}$. The loop phase is therefore $\Phi = \varphi_{1} + \varphi_{2} + \varphi_{3} - \varphi_{1} - \varphi_{2} - \varphi_{3} = 0$ for any directions of the lines of centers, that is, for any shape of the triangle and any twist. It follows from Eq. (\ref{eqS3}), applied to the loop of six modes with the magnitudes $J_{1},J_{2},J_{3},J_{1},J_{2},J_{3}$ (consequence 3 of Section~\ref{secS2}), that the last term of (\ref{eqS3}) is the constant $- 2{J_{1}}^{2}{J_{2}}^{2}{J_{3}}^{2}$, and the determinant does not contain the phases. Explicitly, with $x = \Omega - \omega_{c}$,
\begin{equation*}
\det\left( \Omega\mathbf{I} - \mathbf{H} \right) = \left[ x^{3} - \left( {J_{1}}^{2} + {J_{2}}^{2} + {J_{3}}^{2} \right)x - 2J_{1}J_{2}J_{3} \right]\left[ x^{3} - \left( {J_{1}}^{2} + {J_{2}}^{2} + {J_{3}}^{2} \right)x + 2J_{1}J_{2}J_{3} \right].
\end{equation*}

The two factors are the determinants (\ref{eqS3}) of a loop of three modes with $\Phi = 0$ and with $\Phi = \pi$. For $J_{1} = J_{2} = J_{3} = J$ the local cutoff frequencies are $\omega_{c} + 2J\cos\left( \pi\ell /3 \right)$, $\ell  = 0,\ldots,5$, that is, $\omega_{c} \pm 2J$ and, twice each, $\omega_{c} \pm J$. They do not depend on $z$, so three fibers do not form a lattice.

The cancellation requires all six modes. If the sixth mode is omitted and the fifth mode is coupled directly to the first one, the contact 2--0 is passed only once, its phase $\varphi_{3}$ is not compensated, and the determinant acquires the term $- 2{J_{1}}^{2}{J_{2}}^{2}J_{3}\cos\varphi_{3}$, which is an artifact of the truncation. Likewise, a $3 \times 3$ matrix with one mode per fiber and $\Phi \neq 0$ would describe a bundle in which modes of the same circulation are coupled at one of the contacts, and we neglect this coupling.

\textit{Four fibers under a uniform twist.} Let $N = 4$, $\gamma = \gamma_{0} + \kappa z$ and a constant shape. In the rotating basis and in terms of the supermodes, $\mathbf{P}$ couples $\ell$ and $\ell +2$ with the element $m$, and the dispersion of the axial wavenumber $k_{z}$ is, for $\ell =0,1$,
\begin{equation}\label{eqS12}
\omega_{\ell , \pm}\left( k_{z} \right) = \omega_{c} + D\left( k_{z}^{2} + m^{2}\kappa^{2} \right) \pm 2\sqrt{J^{2}\cos^{2}\left( \pi\ell /2 + \phi \right) + D^{2}m^{2}\kappa^{2}k_{z}^{2}}.
\end{equation}

At $k_{z} = 0$ the cutoff frequencies are $\omega_{c} + \omega_{\kappa} \pm 2J\left| \cos\left( \pi\ell /2 + \phi \right) \right|$ with $\omega_{\kappa} = D(m\kappa)^{2}$, that is, the cutoff frequencies of Eq. (4) of the main text shifted by $\omega_{\kappa}$. With $g = J\left| \cos\left( \pi\ell /2 + \phi \right) \right|$ the minimum of the lower branch over real $k_{z}$ is $\omega_{c} + \omega_{\kappa} - 2g$ for $\omega_{\kappa} \leq g$ and $\omega_{c} - g^{2}/\omega_{\kappa}$ for $\omega_{\kappa} > g$, where it lies at $k_{z} = \pm m\kappa\sqrt{1 - g^{2}/\omega_{\kappa}^{2}}$. The twist does not change the lattice of cutoff frequencies as long as $\omega_{\kappa} < J$, that is, for $\kappa < \sqrt{J/D}/m$, which is 0.24 rad/mm for $J/2\pi = 20$ GHz and $m = 353$. For a larger twist rate the light propagates along helices, which changes the dispersion and not the geometric phase.

\textit{Contrast with twisted multicore fibers.} In Ref. [27] the core modes have no azimuthal structure, so $m_{j} = 0$ and the term (\ref{eqS9}) is absent. The phase of the coupling comes there from the large axial wavenumber $\beta \approx n_{0}k_{0}$ of the core modes, which in the helical frame acts as a transverse momentum proportional to $\beta$ and to the twist rate. A WGM near cutoff has $\beta \rightarrow 0$, so this term vanishes, and the azimuthal momentum $m_{j}$ gives only the redefinition of the phases (\ref{eqS8}). Hence a twist changes the loop phase of the core modes and not that of the WGMs.

\section{Fibers with different cutoff frequencies}\label{secS5}

For four fibers with $\omega_{c,j} = \omega_{c} + \Delta\omega_{c,j}$ and the equal-rate phase $\phi$, Eq. (\ref{eqS7}) with $N = 4$ has the coupling ${\widehat{\Delta}}_{q}$ between supermodes that differ by $q$. The average of the $\Delta\omega_{c,j}$ shifts all branches together. The matrix element between the supermodes $\ell$ and $\ell'$ is $\frac{1}{4}\sum_{j}\Delta\omega_{c,j}e^{i\pi\left( \ell' - \ell  \right)j/2}$, which is
\begin{equation}\label{eqS13}
\begin{gathered}
\frac{1}{4}\left[ \left( \Delta\omega_{c,0} - \Delta\omega_{c,2} \right) \pm i\left( \Delta\omega_{c,1} - \Delta\omega_{c,3} \right) \right]\qquad \left( \ell' - \ell = \pm1 \right), \\
\frac{1}{4}\left( \Delta\omega_{c,0} - \Delta\omega_{c,1} + \Delta\omega_{c,2} - \Delta\omega_{c,3} \right)\qquad \left( \ell' - \ell =2 \right).
\end{gathered}
\end{equation}

\textit{Crossings.} The branches $\ell$ and $\ell +1$ cross where $\cos\left( \phi + \pi\ell /2 \right) = -\sin\left( \phi + \pi\ell /2 \right)$, at the cutoff frequencies $\omega_{c} \pm \sqrt{2}J$, and the branches $\ell$ and $\ell +2$ cross at $\omega_{c}$. For identical fibers these crossings are exact, because the supermodes belong to different eigenvalues $i^{\ell }$ of the cyclic shift. For detuned fibers, degenerate perturbation theory at a crossing gives a gap equal to twice the modulus of the matrix element (\ref{eqS13}):
\begin{equation}\label{eqS14}
\begin{gathered}
\delta_{\omega_{c} \pm \sqrt{2}J} = \frac{1}{2}\sqrt{\left( \Delta\omega_{c,0} - \Delta\omega_{c,2} \right)^{2} + \left( \Delta\omega_{c,1} - \Delta\omega_{c,3} \right)^{2}}, \\
\delta_{\omega_{c}} = \frac{1}{2}\left| \Delta\omega_{c,0} - \Delta\omega_{c,1} + \Delta\omega_{c,2} - \Delta\omega_{c,3} \right|.
\end{gathered}
\end{equation}

These are Eq. (6) of the main text. The first gap is controlled by the mismatch of opposite fibers. If $\Delta\omega_{c,0} = \Delta\omega_{c,2}$ and $\Delta\omega_{c,1} = \Delta\omega_{c,3}$, the half-turn $j \rightarrow j + 2$ remains a symmetry, the parity $( - 1)^{\ell }$ is conserved, and the crossings of supermodes of different parity at $\omega_{c} \pm \sqrt{2}J$ stay exact while those at $\omega_{c}$ open. Away from the crossings the first-order shift vanishes, since ${\widehat{\Delta}}_{0} = 0$ after the average is removed, and the branches move at second order in $\Delta\omega_{c,j}/J$.

\textit{Validity.} It follows from Eq. (\ref{eqS3}) that the four local cutoff frequencies depend on $z$ only through $\Phi$, for any $\Delta\omega_{c,j}$, so each branch repeats with the period of $\Phi$, which is $\Lambda$ for a linear shear, as stated in the main text. The decoupling into Mathieu equations is lost, and the axial dynamics follow Eq. (\ref{eqS4}) with the full matrix. The corrections to the adiabatic potentials are of the order $DK^{2}\left( \widehat{\Delta}/J \right)^{2} = \left( 4J/q_{M} \right)\left( \widehat{\Delta}/J \right)^{2}$ away from the crossings, and they are not small near them, where the gap (\ref{eqS14}) must be passed diabatically, which needs $\Delta\omega_{c,j} \ll J$. For a mismatch comparable with $J$ the structure changes qualitatively, as we describe in the main text and in Section~\ref{secS9}.

\textit{Tolerance.} A change of the fiber radius $\Delta r$ shifts the cutoff frequency by $\omega_{c}\Delta r/r_{0}$, which is $2\pi \times 3.1$ GHz per nanometer for $r_{0} = 62.5$~$\mu$m and $\omega_{c}/2\pi = 193.4$ THz. This agrees with the 30 to 100 GHz shift for 10 to 30 nm radius differences in Ref. [34]. A tolerance of the order of $J/2\pi \approx 20$ GHz corresponds to a radius mismatch of about 6 nm, and a detuning of 0.1 GHz, which is the matching achieved for segments of the same parent fiber [49], to 0.03 nm. The residual mismatch can be trimmed by stretching [41].

\section{Linear shear, tension and contact load}\label{secS6}

Here we consider the case of a four fiber bundle. Let the fiber centers be at $( \pm a,0)$ and $(0, \pm b)$, with $a = 2r_{0}\cos(\theta/2)$ and $b = 2r_{0}\sin(\theta/2)$, so that $a^{2} + b^{2} = 4r_{0}^{2}$ and every side has the length $2r_{0}$. Let $u(z)$ be the transverse coordinate of the axis of a fiber. For small slopes, $|du/dz| \ll 1$, each fiber is longer than a straight fiber by $\frac{1}{2}\int(du/dz)^{2}dz$. With equal axial tensions $T$, the tension energy of the four fibers relative to straight fibers is
\begin{equation}\label{eqS15}
U_{T} = T\int_{0}^{L}\left[ \left( \frac{da}{dz} \right)^{2} + \left( \frac{db}{dz} \right)^{2} \right] dz = Tr_{0}^{2}\int_{0}^{L}\left( \frac{d\theta}{dz} \right)^{2}dz.
\end{equation}

For given end angles $\int(d\theta/dz)dz$ is fixed, and $\int(d\theta/dz)^{2}dz$ is smallest for constant $d\theta/dz$. This is the linear shear $\theta(z) = \pi/2 + \sigma z$. For the symmetric bundle $- L/2 \leq z \leq L/2$ with the end angles $\pi/2 \mp \Delta$, the two end holders are the same device, one rotated by $90^{\circ}$ about the axis.

\textit{Contact load.} Let $u = (a,b)$, let $B = \pi Er_{0}^{4}/4$ be the flexural rigidity of one fiber [43], where $E$ is the Young modulus of silica, and let $p$ be the normal force per unit length at each contact. In the symmetric frictionless model the two neighbors exert the outward loads $pa/r_{0}$ and $pb/r_{0}$ on the fibers on the positive axes, and the small-slope equation of a bent rod under tension [43,44] is $B\, d^{4}u/dz^{4} - T\, d^{2}u/dz^{2} = \left( p/r_{0} \right)u$. For constant $k = (d\theta/dz)/2$ we have $d^{2}u/dz^{2} = - k^{2}u$ and $d^{4}u/dz^{4} = k^{4}u$, so that
\begin{equation}\label{eqS16}
p = r_{0}\left( Tk^{2} + Bk^{4} \right) > 0,
\end{equation}
and adjacent fibers press against each other over the whole span, so that the contacts stay closed. For $E = 72.7$ GPa and $r_{0} = 62.5$~$\mu$m, $B = 8.7 \times 10^{- 7}$ N\,m$^{2}$. For $T = 0.10$ N per fiber and the bundles of Fig. 3 of the main text, sheared over the whole window from $60^{\circ}$ to $120^{\circ}$, Eq. (\ref{eqS16}) gives $p = 0.053$ N/m for $\Lambda = 25$~$\mu$m ($\sigma = 0.178$ rad/mm, $L = 5.9$ mm) and 0.013 N/m for $\Lambda = 50$~$\mu$m ($\sigma = 0.089$ rad/mm, $L = 11.8$ mm). The load integrated along the bundle is about 0.31 and 0.15 mN per contact, respectively.

\textit{Limits of the geometry.} The four side contacts persist for $60^{\circ} \leq \theta \leq 120^{\circ}$. At the ends of this window the two fibers on the shorter diagonal, which are not neighbors, also touch, and a fifth contact appears. Inside the window these two fibers are apart, by 35.7~$\mu$m at $100^{\circ}$ and by 18.4~$\mu$m at $110^{\circ}$. The model takes the centers of touching fibers to be $2r_{0}$ apart in every cross-section $z=\mathrm{const}$. Because the axes of the fibers are inclined to the $z$ axis, a cross-section cuts each fiber in an ellipse, and the centers of two fibers in direct contact are $2r_{0} + r_{0}v_{n}^{2}$ apart, where $v_{n} = \left( r_{0}/2 \right)(d\theta/dz)\sin\theta$ is the slope of their axes along the line of centers. For $\Lambda = 25$~$\mu$m this correction is below 2 nm, a relative change of the side of the rhombus of $1.5 \times 10^{- 5}$, which we neglect. It is not a gap between the fibers and does not change $J$.

\textit{Sensitivity of} $J$ \textit{to the distance between the fibers.} The coupling magnitude depends on the gap $\Delta d$ between the surfaces of adjacent fibers as $J \propto e^{- \sigma_{d}\Delta d}$ with $\sigma_{d} = \omega_{c}\left( n_{0}^{2} - 1 \right)^{1/2}/c \approx 4.2$ $\mu\mathrm{m}^{-1}$ [34], that is, it changes by 0.42\% per nanometer. Since the bottoms of the potential wells of the lattice are at $\omega_{c} - 2J$ (Eq. (4) of the main text), for $J/2\pi = 20$ GHz the eigenfrequencies of the microresonators coincide within a linewidth of 0.1 GHz only if $\Delta d$ is constant along the bundle within 0.6 nm. This cannot be achieved by positioning the fibers, which is why the fibers are assumed to be in direct contact. In contact, $J$ can vary along the bundle only through the deformation of the contacts, which we estimate below.

\textit{Squeezing of the contacts.} A load $p$ per unit length flattens the contact of two parallel fibers over a strip of half-width $\rho = \left( 4pr_{0}/\pi E' \right)^{1/2}$ and brings their centers closer by $\zeta = (4p/\pi E')\left[ \ln\left( 4r_{0}/\rho \right) - 1/2 \right]$ [45]. Here $E' = E/\left( 1 - \nu_{P}^{2} \right) = 74.9$ GPa and $\nu_{P} = 0.17$ is the Poisson ratio of silica. We estimate the change of $J$ as that caused by a change of the gap by $\zeta$. For the loads found above from Eq. (\ref{eqS16}), $p = 0.053$ and 0.013 N/m, we have $\rho = 7.5$ and 3.7 nm and $\zeta = 0.009$ and 0.002 nm, so the tension changes $J$ by less than $10^{- 4}$ of its value. The tolerance of 0.6 nm is reached only at $p \approx 5$ N/m, which is about 100 times larger. Without tension, Eq. (\ref{eqS16}) gives $p = r_{0}Bk^{4}$, which is 0.0034 N/m for $\Lambda = 25$~$\mu$m and 0.0002 N/m for $\Lambda = 50$~$\mu$m. Adjacent fibers then still press against each other, with loads that are 15 and 60 times smaller than for $T = 0.10$ N.

\textit{Fiber ends.} We assume that the holders bring the fibers into contact at the ends of the bundle, for example by elastic fingers. This is necessary because the axes of the fibers are almost straight, departing from straight lines by only a few micrometers over the whole bundle, so that the tension alone presses the fibers together only weakly. A finger that presses a fiber against its neighbors with the force $F$ loads the contacts over a length of about $2L_{c}$ near the finger, where $L_{c} = \left[ 4B/(dp/d\zeta) \right]^{1/4} \approx 0.15$ mm and $dp/d\zeta \approx 8 \times 10^{9}$ N/m$^{2}$ is the stiffness of the contact [44,45]. The largest load is about $F/2L_{c}$, which for $F = 1$ mN gives $\zeta \approx 0.5$ nm and changes $J$ by 0.2\%. This squeezing is confined to the vicinity of the holders.

\textit{Attraction between the fibers.} Silica surfaces attract each other. The nonretarded van der Waals force per unit length between two parallel fibers with the gap $\Delta d \ll r_{0}$ is $p_{W} = A_{H}r_{0}^{1/2}/\left( 16\Delta d^{5/2} \right)$, where $A_{H} \approx 6.5 \times 10^{- 20}$ J is the Hamaker constant of fused silica [47]. This force is 1 N/m for $\Delta d = 1$ nm, 0.065 N/m for 3 nm, where it is equal to the load (\ref{eqS16}) at $\Lambda = 25$~$\mu$m, and $10^{- 5}$ N/m for 100 nm. These values agree with the observations of Ref. [39], in which two fibers intersecting at the angle $\alpha = 1$ mrad stayed in direct contact over about 2 mm. The gap between straight intersecting fibers, $\alpha^{2}z^{2}/4r_{0}$, where $z$ is the distance from the intersection, does not exceed 4 nm over this length. For $\alpha \geq 5$ mrad the attraction had no visible effect [39]. In the bundle the relative tilt of adjacent fibers, $r_{0}\sigma$, is 11 mrad for $\Lambda = 25$~$\mu$m and 5.6 mrad for $\Lambda = 50$~$\mu$m. In contrast to the tilt of two intersecting fibers, it rotates the line of centers and does not open a gap, so the attraction acts along the whole bundle. Electrostatic attraction between charged silica surfaces [47,48] has a longer range and adds to the van der Waals attraction. We do not estimate it here, because it depends on the surface charge of the fibers.

Fibers in contact adhere to each other. For a work of adhesion of 0.05 J/m$^{2}$, which we assume for silica surfaces in air, the adhesive contact of two parallel fibers has a half-width of about 0.2~$\mu$m at zero load [46] and brings their centers closer by less than about 1 nm. An adhesion that is uniform along the bundle changes $J$ by the same amount in all cross-sections and does not disturb the lattice. The adhesion energy per unit length of the contact is about $2 \times 10^{- 8}$ J/m. A contact that is closed at one point advances along the fibers as long as the curvature that the fibers need to close the gap ahead of it is smaller than about 0.2 m$^{-1}$, at which the bending energy per unit length of a fiber [43] is equal to this adhesion energy. This curvature closes a gap of 0.1~$\mu$m over a length of 1 mm. The attraction therefore closes the residual gaps between fibers that are in contact at the ends and keeps the contacts closed, and the tension adds the load (\ref{eqS16}).

\section{Microresonator lattice: parameters, eigenfrequencies and tunability}\label{secS7}

\textit{Parameters.} We use $J/2\pi = 20$ GHz, the numerical value for contacting 62.5~$\mu$m radius silica fibers in Ref. [34], where the potential well depth of two coupled fibers is 19.7 GHz, $n_{0} = 1.44$ and the wavelength 1.55~$\mu$m, which give $D/2\pi = 2843$ GHz $\mu\mathrm{m}^{2}$, and $m = 353$. The full depth $4J$ of the cosine potential of Eq. (4) of the main text corresponds to an effective radius variation $\Delta r_{\mathrm{eff}} = 4Jr_{0}/\omega_{c}$ of about 26 nm. For the linear shear the Mathieu parameter is $q_{M} = 4J/DK^{2}$ with $K = m\sigma = \pi/(2\Lambda)$, where $\Lambda$ is the distance between neighboring microresonators (Eq. (5) of the main text) and $4\Lambda$ is the period of the lattice of each supermode. For $q_{M} \gg 1$ each potential well supports several localized states and the bands are narrow, and for $q_{M} \lesssim 1$ the spectrum is nearly free with small gaps. With the numbers above, $q_{M} = 456$, 114, 28.5 and 7.1 for $\Lambda = 200$, 100, 50 and 25~$\mu$m, and $q_{M} = 1$ at $\Lambda \approx 9.4$~$\mu$m. For the weak coupling $J/2\pi \approx 1$ GHz, which corresponds to a gap of 0.8~$\mu$m between the fibers [34], $q_{M} \approx 0.4$ at $\Lambda = 25$~$\mu$m and no resonators form.

\textit{Eigenfrequencies and modes of Fig. 3.} We solved Eq. (4) of the main text by finite differences with a step of 0.25~$\mu$m for a closed lattice of nine periods $4\Lambda$. The eigenfrequencies fall into bands with one state per potential well. For $\Lambda = 25$~$\mu$m there are three bands below the barrier top, centered at $- 25.77$, $1.00$ and $23.8$ GHz relative to $\omega_{c}$, with the widths 0.006, 0.20 and 2.2 GHz. For $\Lambda = 50$~$\mu$m there are seven, centered at $- 32.69$, $- 18.46$, $- 5.03$, $7.52$, $19.04$, $29.28$ and $37.8$ GHz, with the widths below $10^{- 4}$ GHz for the two lowest and 0.0003, 0.006, 0.07, 0.60 and 2.7 GHz for the others. The shaded profiles of Fig. 3 are the intensities summed over the states of a band and shown within one potential well. The intensity of the lowest mode exceeds half of its maximum over 23.5~$\mu$m for $\Lambda = 25$~$\mu$m and over 32.5~$\mu$m for $\Lambda = 50$~$\mu$m. The number of bands that lie entirely below the barrier top is 27 at $\Lambda = 200$~$\mu$m, 13 at 100~$\mu$m, 7 at 50~$\mu$m and 3 at 25~$\mu$m.

\textit{Smallest period with isolated microresonators of a lattice.} The width of the lowest band follows from the characteristic values of the Mathieu equation [42]. It is 0.0005 GHz at $\Lambda = 31$~$\mu$m, 0.0065 GHz at 25~$\mu$m, 0.06 GHz at 20~$\mu$m, 0.10 GHz at 19~$\mu$m, 0.56 GHz at 15~$\mu$m and 1.7 GHz at 12.5~$\mu$m. Light excited in one potential well stays in it if this width is smaller than the resonance linewidth. For a linewidth of 0.1 GHz this requires $\Lambda \gtrsim 19$~$\mu$m. This condition concerns the microresonators of one lattice, which are $4\Lambda$ apart. Neighboring microresonators of different lattices, $\Lambda$ apart, overlap at this spacing and are uncoupled for identical fibers only (Section~\ref{secS9}). The lowest eigenfrequency lies below the lowest crossings of the cutoff frequencies, at $- \sqrt{2}J$, for $\Lambda \geq 31$~$\mu$m. A band wider than the linewidth describes a chain of coupled microresonators with coinciding eigenfrequencies. In the tight-binding approximation a band of width $W$ gives a delay of about $1/\pi W$ per microresonator at the band center, which is about 3 ns for the 20 microresonators of the third band at $\Lambda = 25$~$\mu$m ($W = 2.3$ GHz for 20 potential wells, length 2 mm). We did not calculate the transmission of such a delay line.

\textit{Spacing and length.} The spacing, $\Lambda = \pi/(2m|\sigma|)$, can be changed by the relative angle of the end holders or their axial translation. One period $4\Lambda$ of a lattice requires a change of the rhombus angle by $2\pi/m$, so a sheared section with $N_{r}$ microresonators has the length $N_{r}\Lambda$ and requires a change of the rhombus angle by $\Delta\theta = \pi N_{r}/(2m)$. The window from $60^{\circ}$ to $120^{\circ}$ allows at most $m/6 \approx 59$ periods of each lattice, that is, $2m/3 \approx 235$ microresonators, for any $\Lambda$, and a bundle sheared over the whole window has the length
\begin{equation}\label{eqS17}
L = \frac{2m\Lambda}{3},
\end{equation}
which is 5.9 mm for $\Lambda = 25$~$\mu$m, 11.8 mm for 50~$\mu$m and 58.8 mm for 250~$\mu$m. A span $L$ fixes the smallest spacing, $\Lambda_{\min} = 3L/(2m)$. For end angles that differ by $\Delta\theta$ the length is $2m\Lambda\Delta\theta/\pi$. Table~\ref{tabS1} lists design points.

\begin{figure}[tb]
\centering\includegraphics[width=0.75\textwidth]{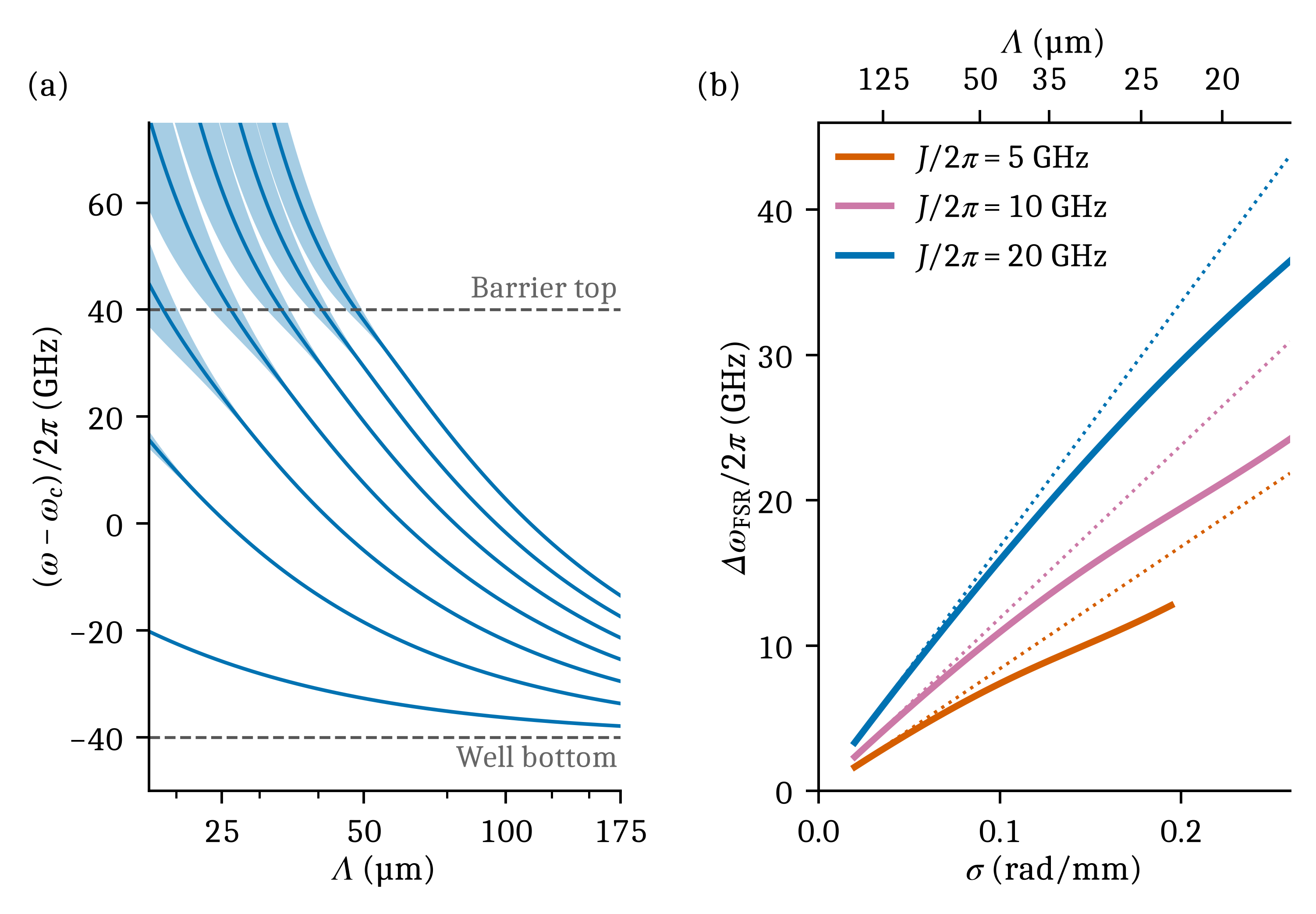}
\caption{\label{figS1}Tunability of the lattice for $m = 353$. (a) Allowed bands of the Mathieu equation (shaded, with band centers as lines) versus $\Lambda$ for $J/2\pi = 20$ GHz. Dashed lines mark the potential well bottom and the barrier top at $\mp 2J$. (b) Spacing between the two lowest bands versus gradient for $J/2\pi = 5$, 10 and 20 GHz (solid), with the harmonic estimate, Eq. (\ref{eqS18}), dotted. Curves end where the second band reaches the barrier top.}
\end{figure}

\textit{Spacing of the eigenfrequencies.} For $q_{M} \gg 1$ the lowest eigenfrequencies of each potential well are those of a harmonic oscillator with the potential $\omega_{c} - 2J + JK^{2}z^{2}$, with the spacing
\begin{equation}\label{eqS18}
\Delta\omega_{\mathrm{FSR}} \approx 2K\sqrt{DJ} = 2m\sigma\sqrt{DJ},
\end{equation}
which is linear in the gradient $\sigma$, as the free spectral range of two crossing fibers is linear in their tilt [34]. Eq. (\ref{eqS18}) gives $\Delta\omega_{\mathrm{FSR}}/2\pi =3.7$, 7.5, 15.0 and 30.0 GHz for the four design points of Table~\ref{tabS1}, within 12\% of the exact values. At a fixed gradient the spacing grows as $\sqrt{J}$. Figure~\ref{figS1} shows the bands and the spacing of the eigenfrequencies as functions of $\Lambda$ and of the gradient.

\textit{Position of the lattice.} A change $\delta\theta$ of the angle at the ends moves every potential well by $\delta z = \delta\theta/\sigma$ and changes the loop phase by $\delta\Phi = 4m\,\delta\theta$. A closed-loop piezoelectric stage with 10 nm repeatability [34] that moves one finger of a holder changes $\theta$ by $a\,\delta b/\left( 2r_{0}^{2} \right) \approx 0.1$ mrad, which shifts the lattice by 0.6~$\mu$m at $\Lambda = 25$~$\mu$m. The spacing $\Lambda$ depends on the gradient only and is insensitive to the absolute position of the fingers.

\begin{table}[htb]
\caption{\label{tabS1}Design points for $m = 353$, $J/2\pi = 20$ GHz. $r_{0}\sigma$ is the tilt of adjacent fibers relative to each other. The last column is the length $N_{r}\Lambda$ of a sheared section with $N_{r} = 100$ microresonators.}
\centering\small
\begin{tabular}{cccccc}
\toprule
$\Lambda$ ($\mu$m) & $\sigma$ (rad/mm) & $r_{0}\sigma$ (mrad) & $q_{M}$ & \shortstack{Spacing of the two\\ lowest bands (GHz)} & \shortstack{Length for\\ $N_{r}=100$ (mm)} \\
\midrule
200 & 0.022 & 1.4 & 456 & 3.7 & 20 \\
100 & 0.044 & 2.8 & 114 & 7.3 & 10 \\
50 & 0.089 & 5.6 & 28.5 & 14.2 & 5 \\
25 & 0.178 & 11.1 & 7.1 & 26.8 & 2.5 \\
\bottomrule
\end{tabular}
\end{table}

\textit{Lattices of finite length.} In Fig. 2 of the main text the bundle is sheared over the length $L$ between the two holders. The holders touch the fibers and introduce a strong loss, so that the WGMs are absorbed there. We model this by a loss that rises over 100~$\mu$m to 300 GHz inside the holders, which is an assumed model of the loss at the holders. For a sheared section with $N_{r}$ microresonators, we chose the angles at the holders at which the cutoff frequency of the supermode $\ell =2$ has its maximum, so that this supermode has $N_{r}/4$ identical potential wells between the holders. For $\Lambda = 25$~$\mu$m and $N_{r} = 80$ (a sheared length of 2 mm and a change of the rhombus angle by ${20.4}^{\circ}$) the three bands lie at $- 25.77$ GHz, from 0.90 to 1.11 GHz and from 22.7 to 25.0 GHz, and the band positions are the same for $N_{r} = 20$. For $N_{r} = 80$ the loss at the holders adds less than 0.001 GHz to the linewidths of the states of the two lower bands and up to 0.004 GHz to those of the third band. For $\Lambda = 50$~$\mu$m and $N_{r} = 40$ we find the seven bands listed above.

\section{Spectrogram measured with a transverse microfiber}\label{secS8}

\begin{figure}[tb]
\centering\includegraphics[width=0.72\textwidth]{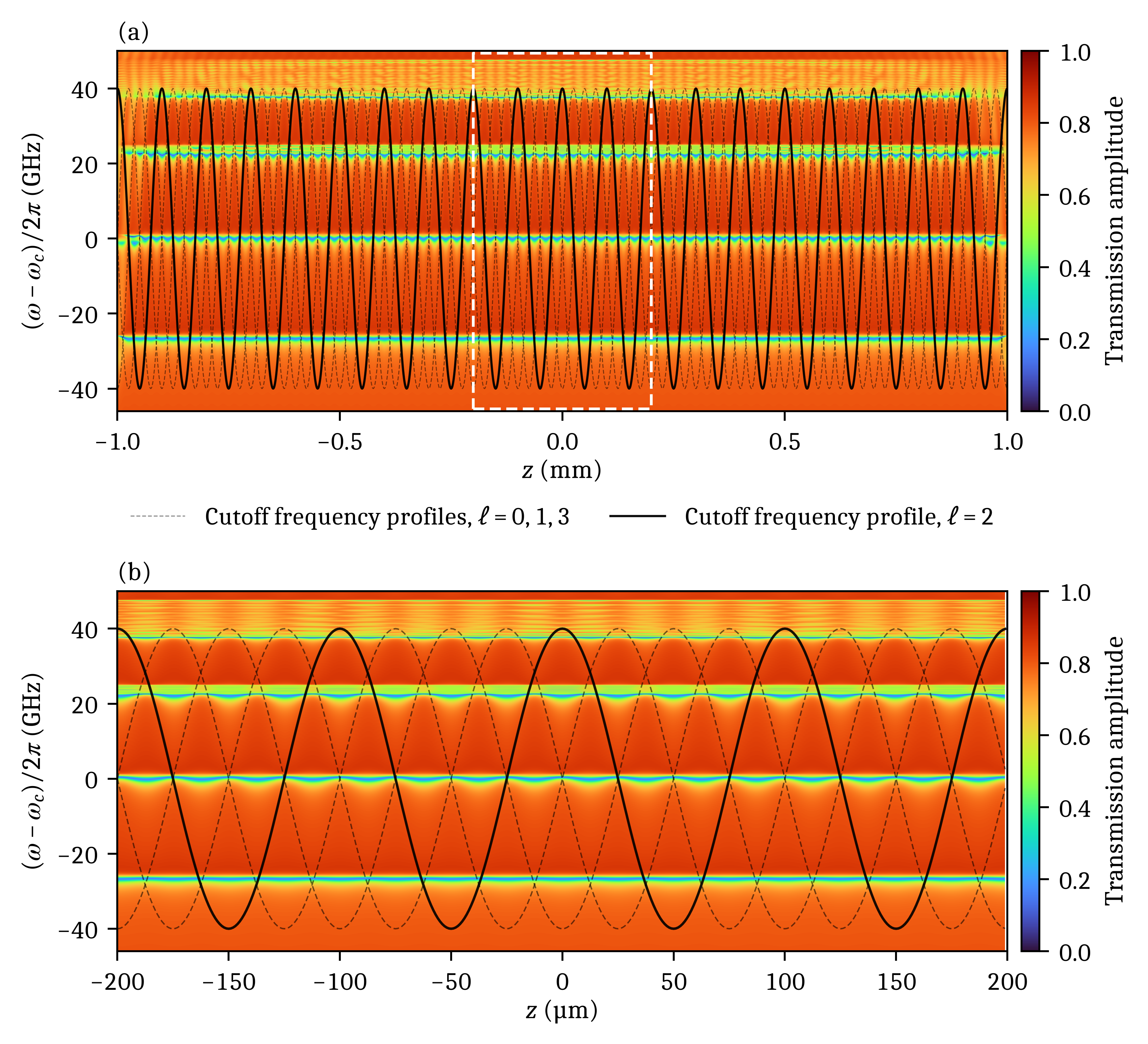}
\caption{\label{figS2}Calculated spectrogram of a bundle of four identical fibers with $N_{r} = 80$ microresonators spaced by $\Lambda = 25$~$\mu$m, probed by a transverse microfiber in contact with one of the fibers; $J/2\pi = 20$ GHz, intrinsic linewidth 0.1 GHz, the microfiber coupling parameters of Ref. [35], and absorbing holders at the ends of the sheared section. (a) Transmission amplitude versus the microfiber position and frequency over the whole sheared section. Black lines are the cutoff frequency profiles of the supermode $\ell =2$ (solid) and of the other three supermodes (dashed). (b) The central part of the bundle with 16 microresonators. The resonances repeat with the period $\Lambda$ because the microfiber excites all four lattices.}
\end{figure}

Figure~\ref{figS2} shows what a transverse microfiber scanned along a bundle of four identical fibers would measure. We calculated the transmission amplitude $S = S_{0} - i|C|^{2}G/\left( 1 + D_{c}G \right)$ of Ref. [35], where $G$ is the Green's function of Eq. (2) of the main text at the microfiber position, for the sheared section between the absorbing holders described in Section~\ref{secS7}. We used the coupling parameters found experimentally for a chain of SNAP microresonators in Ref. [35], $S_{0} = 0.83 - 0.15i$, $|C|^{2} = 0.028$ $\mu\mathrm{m}^{-1}$ and $D_{c} = 0.022 + 0.026i$ $\mu\mathrm{m}^{-1}$, and an intrinsic linewidth of 0.1 GHz. These parameters are an assumption for the bundle.

A microfiber that touches one fiber excites the four supermodes equally, because each of them has the same intensity in every fiber. The spectrogram therefore contains all four lattices, and the resonances of each band repeat along the bundle with the period $\Lambda$, although each lattice has the period $4\Lambda$.

\textit{Excitation of a single lattice or a single microresonator.} The four lattices have identical spectra and cannot be separated by frequency. However, they can be separated by phase. Two microfibers in contact with the opposite fibers 0 and 2 select the even supermodes when they are driven in phase and the odd supermodes when they are driven in antiphase, and four inputs with the phase step $\pi\ell /2$ select the supermode $\ell$. Alternatively, the bundle can be made to contain one microresonator only. If the rhombus angle changes between the holders by $\pi/(2m)$ only, which is ${0.25}^{\circ}$ for $m = 353$, the sheared section has the length $L = \Lambda$, and the phase satisfies $|\phi| < \pi/4$ along the whole of it. The supermode $\ell =2$, whose cutoff frequency has its minimum at the center of the bundle, where the cross-section is a square, then has the lowest cutoff frequency of the four supermodes everywhere between the holders. Its cutoff frequency changes from $\omega_{c} - \sqrt{2}J$ at the holders to $\omega_{c} - 2J$ at the center, a single potential well of depth $\left( 2 - \sqrt{2} \right)J \approx 0.59J$, and the states of this well lie below the cutoff frequencies of all four supermodes along the whole sheared section. It follows from Eq. (\ref{eqS18}) with $K = \pi/(2L)$ that the lowest eigenfrequencies of the well are spaced by $(\pi/L)\sqrt{DJ}$, which is 0.37 GHz for $L = 2$ mm. The lowest states are localized near the center of the bundle, away from the holders, where the WGMs are absorbed.

\section{One detuned fiber and the coupling between neighboring microresonators}\label{secS9}

\textit{Detuning comparable with} $J$\textit{.} We solved Eq. (2) of the main text with the diagonal $\mathrm{diag}\left( \Delta\omega_{c,j} \right)$ directly for the four coupled envelopes, without the adiabatic approximation. With one fiber detuned by $\Delta\omega_{c,0}/2\pi = 50$ GHz and the other three equal, the four crossing cosines of depth $4J/2\pi = 80$ GHz are replaced by four separated branches with the modulation depths 5.2, 12.8, 9.9 and 2.4 GHz and the period $\Lambda$. The highest branch lies near the cutoff frequency of the detuned fiber and carries about 80\% of its intensity in that fiber. With a detuning of $\Delta\omega_{c,0}/2\pi = 20$ GHz the depths are 8.0, 20.0, 18.9 and 6.9 GHz. Since these depths are much smaller than $4J$ and the potential wells are four times narrower than for identical fibers, the spectrograms we calculated as in Section~\ref{secS8} show no lattice of separate microresonators at $\Lambda = 25$ and 50~$\mu$m, where the resonances extend along the whole sheared section as bands with a weak modulation of period $\Lambda$. At $\Lambda = 125$~$\mu$m the two deepest branches hold resonances in individual potential wells, and at $\Lambda = 250$~$\mu$m all four branches do. In contrast with identical fibers, the spectrogram depends on which fiber the microfiber touches, since the branches are distributed unequally among the fibers.

\begin{figure}[tb]
\centering\includegraphics[width=0.68\textwidth]{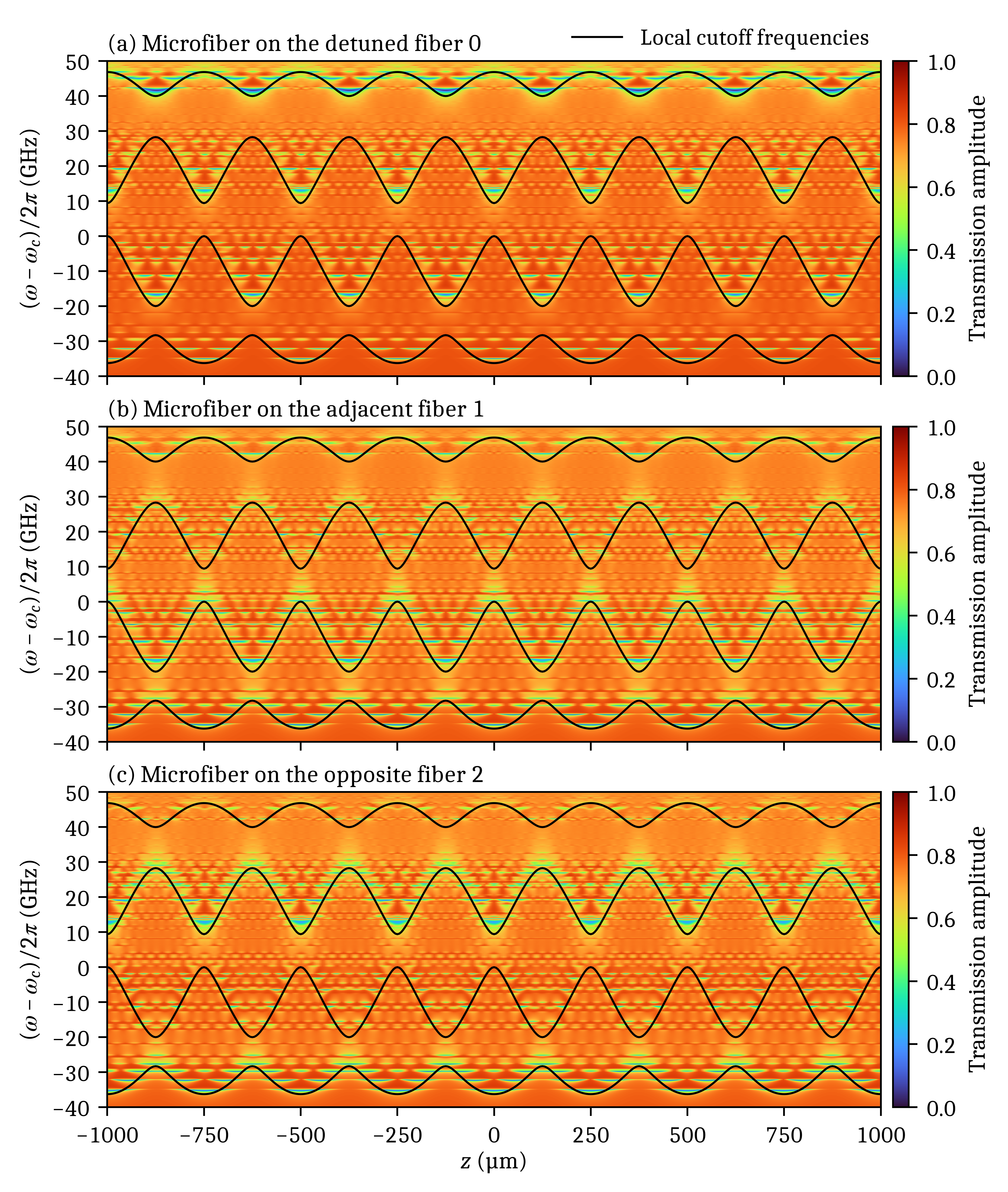}
\caption{\label{figS3}Calculated spectrogram of the bundle of Fig. 4 of the main text: one fiber detuned by $\Delta\omega_{c,0}/2\pi = + 20$ GHz, 40 periods of $\Lambda = 250$~$\mu$m, $J/2\pi = 20$ GHz, absorbing ends. The eight central periods are shown. The transverse microfiber is in contact with (a) the detuned fiber 0, (b) the adjacent fiber 1 and (c) the opposite fiber 2. The intrinsic linewidth is 0.1 GHz, and the microfiber coupling parameters are those of Ref. [35], as in Fig.~\ref{figS2}. Black lines are the four local cutoff frequencies.}
\end{figure}

\textit{Localized states of Fig. 4.} Figure 4 of the main text is calculated for a detuning of $\Delta\omega_{c,0}/2\pi = 20$ GHz and $\Lambda = 250$~$\mu$m with a grid step of 2.5~$\mu$m. The ends are made absorbing by a loss that rises over 100~$\mu$m to 300 GHz inside the holders, which is an assumed model of the loss at the holders. Figure 4 shows the states that are confined to the potential wells and do not reach the holders. In the calculation we selected them as the states whose broadening by the loss at the holders is much smaller than the intrinsic linewidth of 0.1 GHz assumed in Fig.~\ref{figS3} (we used a limit of 0.01 GHz). The lowest branch, whose potential wells are 8 GHz deep and 250~$\mu$m apart, supports three bands with one state per well, at $- 34.85$, $- 32.05$ and $- 29.4$ GHz, with the widths 0.007, 0.10 and 0.60 GHz. The upper branches support several eigenfrequencies per well. Their states have the same frequencies as extended states of the lower branches. In our calculation a part of them mixes with these extended states and reaches the absorbing ends. Their linewidths are estimated below.

\textit{Spectrogram of the bundle of Fig. 4.} Figure~\ref{figS3} shows the spectrogram of the bundle of Fig. 4 of the main text, calculated as in Section~\ref{secS8} with a grid step of 1~$\mu$m for a microfiber in contact with the detuned fiber 0, with the adjacent fiber 1 and with the opposite fiber 2. The lowest band is at $-34.84$ GHz, compared with $-34.85$ GHz in Fig. 4. The resonances lie in the potential wells of the four branches of the local cutoff frequency, and their depth in the spectrogram depends on which fiber the microfiber touches. The lowest state of the lowest branch has 14\% of its intensity in fiber 0, 27\% in each of the fibers 1 and 3, and 33\% in fiber 2, so it is seen through any fiber. The resonance of the highest branch at 42.5 GHz has 65\% of its intensity in fiber 0, 17\% in each of the fibers 1 and 3, and 1\% in fiber 2, so the highest branch is seen mainly through the detuned fiber and is absent in Fig.~\ref{figS3}(c).

\textit{Overlap of the microresonators.} The microresonators of this bundle overlap in three ways. All three are contained in the solution of Eq. (2) of the main text, which we obtain without the adiabatic approximation. First, microresonators of different branches overlap in space. The potential wells of the first and third branches are centered at the same positions, $z = n\Lambda$ from the center of the bundle, and the wells of the second and fourth branches lie halfway between them, at a distance $\Lambda/2=125$~$\mu$m. Each section of the bundle therefore holds microresonators of two branches, whose frequencies differ by more than 37 GHz, and a microfiber at any position excites both of them.

Second, neighboring microresonators of the same branch overlap when their eigenfrequency is close to the top of the barrier. At the lowest eigenfrequency, $-34.84$ GHz, the response (the imaginary part of the Green's function summed over the four fibers) in the middle of a barrier is 0.001 of the response at the center of a potential well, and the microresonators are separate. At the third band of the lowest branch, near $-29.4$ GHz, which is 0.60 GHz wide, this ratio is 0.64, and the resonance is continuous along the bundle.

Third, the eigenfrequency of every microresonator of the three upper branches lies above the barriers of the branches below it. At this frequency the lower branches have states that are not confined to the potential wells and extend along the whole bundle to the holders. For unequal fibers the branches are weakly coupled by the variation of the phase along the bundle, so that a microresonator of an upper branch leaks into these extended states. To estimate the linewidth that this adds, we recalculated the spectra at the centers of the two potential wells nearest to the center of the bundle ($z = 0$ and 125~$\mu$m) with the intrinsic linewidth reduced to 0.002 GHz and a frequency step of 0.0005 GHz, in intervals of 1.2 GHz around the 30 strongest resonances. Most resonances then resolve into groups of narrow lines, which are a few hundredths to a few tenths of a GHz apart. In total we found 366 lines. Their widths exceed the intrinsic 0.002 GHz by a median of 0.001 GHz below $-28.3$ GHz, where only the lowest branch exists (largest value 0.009 GHz), by 0.005 GHz between $-28.3$ and 0 GHz (largest 0.07 GHz), by 0.007 GHz between 0 and 28.3 GHz (largest 0.09 GHz), and by 0.014 GHz above 28.3 GHz (largest 0.12 GHz). Some resonances stay single, for example those at $-6.34$ GHz and at 42.49 GHz, whose widths exceed the intrinsic one by less than 0.001 GHz.

In this model the loss through the extended states to the absorbing ends therefore adds about 0.01 GHz, and at most about 0.1 GHz, to the linewidth of a microresonator of the upper branches, which is not more than the intrinsic linewidth of 0.1 GHz assumed in Fig.~\ref{figS3}. These values are found for a bundle of 40 periods and for the assumed loss at the holders, and we have not determined how they change with the length of the bundle. Within a group of lines we did not separate the splitting caused by the coupling of neighboring potential wells from the splitting caused by the extended states.

\textit{Neighboring microresonators have equal eigenfrequencies for any uniform detuning.} A shift $z \rightarrow z + \Lambda$ changes the phase in Eq. (2) of the main text by $\pi/2$. The constant redefinition of the phases
\begin{equation}\label{eqS19}
\psi_{j} \rightarrow e^{- i\pi j/2}\psi_{j}
\end{equation}
removes this change, because it multiplies the coupling of the modes $j$ and $j + 1$ by $e^{- i\pi/2}$ and is single valued around the loop. It does not affect the diagonal $\mathrm{diag}\left( \Delta\omega_{c,j} \right)$. Hence Eq. (2) with any cutoff frequencies that are constant along the fibers is invariant under a shift by $\Lambda$, and all potential wells, which are $\Lambda$ apart, have the same eigenfrequencies. A detuning that is uniform along the bundle cannot separate the microresonators in frequency. To select a single microresonator by frequency the cutoff frequency must vary along the bundle, for example through a nonlinear change of the rhombus angle or a slow change of the radius of one fiber.

\textit{Coupling between neighboring microresonators.} For identical fibers the microresonators in neighboring potential wells are formed by different supermodes, and their eigenmodes are not coupled. A detuning $\Delta\omega_{c,0}$ of one fiber couples the supermodes with the matrix element $\Delta\omega_{c,0}/4$, Eq. (\ref{eqS13}). For $\Delta\omega_{c,0} \ll J$ the states of two neighboring wells are then coupled with the strength $t = \Delta\omega_{c,0}\xi/4$, where $\xi$ is the overlap of their axial envelopes, their common eigenfrequency shifts by $\Delta\omega_{c,0}/4$, and in the nearest-neighbor approximation it becomes a band of width $4t = \Delta\omega_{c,0}\xi$. For Gaussian envelopes of the lowest state, $\exp\left( - z^{2}/2w^{2} \right)$ with $w = (D/J)^{1/4}K^{- 1/2}$, the overlap of potential wells $\Lambda$ apart is $\xi =\exp\left( - \Lambda^{2}/4w^{2} \right)$, which is 0.44 for $\Lambda = 25$~$\mu$m ($w = 13.8$~$\mu$m) and 0.19 for $\Lambda = 50$~$\mu$m ($w = 19.5$~$\mu$m). We checked this estimate by solving Eq. (2) by finite differences with a step of 0.25~$\mu$m for a closed lattice of five periods $4\Lambda$ with one fiber detuned by $\Delta\omega_{c,0}/2\pi = 0.25$ to 4 GHz. The width of the lowest band is $0.48\Delta\omega_{c,0}$ at $\Lambda = 25$~$\mu$m (0.48 GHz at 1 GHz and 1.88 GHz at 4 GHz) and $0.21\Delta\omega_{c,0}$ at $\Lambda = 50$~$\mu$m (0.21 GHz at 1 GHz and 0.82 GHz at 4 GHz), and its center shifts by $\Delta\omega_{c,0}/4$.

\textit{Consequences.} Stretching one fiber tunes the coupling between neighboring microresonators continuously from zero. As long as the band is narrower than the resonance linewidth, a microfiber excites the single microresonator at its position. For a wider band, light of a given frequency excites an eigenmode of the whole chain of microresonators. The eigenmodes of a chain of $n$ microresonators are separated by about the band width divided by $n$, which sets the Q-factor needed to excite one of them.

\textit{Control by stretching.} Axial stretching of a fiber shifts its cutoff frequency. According to Ref. [41], a relative elongation of about $10^{- 3}$ gives a tuning range of about 100 GHz, which covers the whole interval from $\Delta\omega_{c,0} \ll J$ to $\Delta\omega_{c,0} \gg J$. Stretching one fiber therefore changes the structure continuously from four interleaved lattices of period $4\Lambda$ and depth $4J$ to a lattice of period $\Lambda$ whose depth and distribution among the fibers are set by the detuning.

\section{Scope of the model}\label{secS10}

We used the nearest-neighbor coupling between the dominant counter-circulating WGMs of one radial family per fiber. We neglected other WGM families that contribute to the cutoff shift at small gaps [34], elastic flattening at the contacts, which is estimated in Section~\ref{secS6}, and scattering at the fiber surfaces. Two further effects break the cyclic symmetry and open the crossings of the cutoff frequencies in the same way as a mismatch of the fibers: unequal coupling magnitudes on the four sides, and the coupling across the two diagonals of the rhombus, which have different lengths. The latter is exponentially small in the diagonal gaps and leaves the crossings at $\omega_{c} \pm \sqrt{2}J$ intact, but it should be estimated for angles close to $60^{\circ}$ and $120^{\circ}$. Residual axial variations of the radius of each fiber add shifts $\Delta\omega_{c,j}(z)$ of the cutoff frequencies, which must be small compared with $4J$ to keep the lattice periodic. We expect the WGMs to be absorbed where the fibers touch the holders. We included this loss in the calculated spectrograms and in Fig. 4 of the main text by an assumed model (Sections~\ref{secS7} and \ref{secS9}). In these calculations it adds less than 0.004 GHz to the linewidths of the states localized in the potential wells for identical fibers (Section~\ref{secS7}), and about 0.01 GHz, at most about 0.1 GHz, for the bundle with one detuned fiber (Section~\ref{secS9}), compared with the assumed intrinsic linewidth of 0.1 GHz. A twist of the holders adds a constant to all cutoff frequencies and does not change the period (Section~\ref{secS4}).

\end{document}